\documentclass[useAMS,usenatbib]{mn2e}

\usepackage{tabularx}
\usepackage{graphicx}
\usepackage[section]{placeins}
\usepackage{epstopdf}
\usepackage{array}
\usepackage[fleqn]{amsmath}
\usepackage{amssymb}
\usepackage{wrapfig}
\usepackage{upgreek}
\usepackage{color}
\usepackage{longtable}
\usepackage{supertabular}
\usepackage{units}
\usepackage{booktabs}
\usepackage{multirow}

\def \co12 {$^{12}$CO}
\def \co13 {$^{13}$CO}
\def \co {C$^{18}$O}
\def \nh {N$_{2}$H$^{+}$}
\def \sun {$_{\odot}$}

\def \mnras {MNRAS}
\def \aap {A\&A}

\def \pasp {PASP}
\def \apj {ApJ}

\def \apjl {ApJL}

\def \nature {Nature}
\def \araa {ARA\&A}

\DeclareMathOperator\erf{erf}

\title[An Analytical Model for Starless Cores]{An analytical model for the evolution of starless cores~I: The constant-mass case}
\author[K. Pattle]{K. Pattle\\
\\
Jeremiah Horrocks Institute, University of Central Lancashire, Preston PR1 2HE, United Kingdom}

\begin{document}

\date{}

\pagerange{\pageref{firstpage}--\pageref{lastpage}} \pubyear{2016}

\maketitle

\label{firstpage}

\begin{abstract}

We propose an analytical model for the quasistatic evolution of starless cores confined by a constant external pressure, assuming that cores are isothermal and obey a spherically-symmetric density distribution.  We model core evolution for Plummer-like and Gaussian density distributions in the adiabatic and isothermal limits, assuming Larson-like dissipation of turbulence.  We model the variation in the terms in the virial equation as a function of core characteristic radius, and determine whether cores are evolving toward virial equilibrium or gravitational collapse.  We ignore accretion onto cores in the current study.  We discuss the different behaviours predicted by the isothermal and adiabatic cases, and by our choice of index for the size-linewidth relation, and suggest a means of parameterising the magnetic energy term in the virial equation.  We model the evolution of the set of cores observed by \citet{pattle2015} in the L1688 region of Ophiuchus in the `virial plane'.  We find that not all virially-bound and pressure-confined cores will evolve to become gravitationally bound, with many instead contracting to virial equilibrium with their surroundings, and find an absence of gravitationally-dominated and virially-unbound cores.  We hypothesise a `starless core desert' in this quadrant of the virial plane, which may result from cores initially forming as pressure-confined objects.  We conclude that a virially-bound and pressure-confined core will not necessarily evolve to become gravitationally bound, and thus cannot be considered prestellar.  A core can only be definitively considered prestellar (collapsing to form an individual stellar system) if it is gravitationally unstable.

\end{abstract}


\section{Introduction}

Starless cores are the immediate precursors to the formation of protostars: small-scale overdensities within a molecular cloud which will, if gravitationally unstable, collapse to form an individual star or system of stars \citep{beichman1986}.  Not all starless cores will go on to form protostellar systems.  The gravitationally-unstable and collapsing subset of starless cores are known as prestellar cores \citep{wardthompson1994}.  Barring external disruption, an individual stellar system will form from a prestellar core.  Understanding the properties of starless cores is essential to understanding the stars which will one day form from them; recent studies have shown an apparent link between the core mass function and the intial mass function (see \citealt{offner2014}, and references therein).

There is a paucity of analytic and semi-analytic evolutionary models for starless cores.  Historically, the Singular Isothermal Sphere model (\citealt{shu1977}; \citealt{shu1987}) has been used.  However, the dynamical instability of this model was noted by \citet{whitworth1996}, and more recent high-resolution observations have shown that starless cores have a non-singular geometry, typically with a flat central plateau (e.g. \citealt{alves2001}).  Frequently-used starless core geometries include the Plummer density distribution \citep{plummer1911}, which was first applied to starless cores by \citet{whitworth2001}, and the Bonnor-Ebert (BE) density distribution (\citealt{ebert1955}; \citealt{bonnor1956}).  The BE density distribution is parameterised by its central density and characterised by a plateau of slowly decreasing density at small radii and a power-law decrease at large radii, and which treats a core as an isothermal, self-gravitating, polytropic sphere bounded by external pressure.  At least some starless cores appear to be well-characterised by a BE density distribution \citep{alves2001}, and the BE critically-stable mass of a starless core is often treated as a proxy for virial mass (e.g. \citealt{konyves2015}).  However, the BE model requires the Lane-Emden equation to be solved numerically, and so modelling the evolution of a Bonnor-Ebert sphere remains the preserve of computational astrophysics (e.g. \citealt{broderick2010}; \citealt{keto2015}).

Starless cores have been shown to be confined by external pressure in many cases (e.g. \citealt{alves2001}; \citealt{johnstone2000}; \citealt{maruta2010}).  \citet{pattle2015} found a population of virially-bound starless cores in the Ophiuchus molecular cloud, for which external pressure significantly dominates over gravity in core confinement.  It is not clear whether a virially bound and pressure-confined starless core will evolve to become gravitationally unstable.  Scenarios can be envisaged in which a core will contract under pressure until it becomes self-gravitating, or alternatively, where the core will contract to virial equilibrium with its surroundings.

In this paper we construct a model for the quasistatic evolution of a spherically-symmetric pressure-confined starless core.  We consider two fully analytic solutions to this model: (1) a truncated Plummer-like density distribution (which produces a density profile similar to that of a Bonnor-Ebert distribution) and (2) a truncated Gaussian density distribution.  The model is intended to assess whether an observed starless core is likely to evolve to become gravitationally unstable, or to virial equilibrium with its surroundings.  We model cores in the non-magnetic and magnetised cases.  The non-magnetic model has six initial parameter conditions, all of which are observable quantities: mass, size, temperature, internal velocity dispersion, external velocity dispersion and external density.  The magnetised case introduces a seventh observable initial condition, magnetic field strength.  In this paper, we assume negligible accretion of mass onto the core.  In a subsequent paper, we will consider the case in which a core can continue to accrete mass.

The model presented in this paper is envisaged as a means by which the likely evolutionary outcome -- gravitational collapse, or virial equilibrium -- of an observed starless core can be rapidly assessed, without the need for computationally expensive simulations to be performed.

In Section 2 we formulate the model.  In Section 3 we consider Plummer-like and Gaussian density profiles.  In Section 4 we discuss core stability as a function of radius and construct evolutionary tracks in the virial plane.  In Section 5 we discuss the cases of adiabatic and isothermal core contraction.  In Section 6 we discuss the parameterisation of non-thermal motions.  In Section 7 we discuss the parameterisation of the magnetic field.  In Section 8 we apply this model to the starless cores identified in the L1688 region of the Ophiuchus molecular cloud by \citet{pattle2015}.  In Section 9 we discuss our results, and in Section 10 we summarise our conclusions.

\section{Formulation of the Evolutionary Model}

We first present the general case of the model, without specifying a core geometry.

\subsection{Core density profile}
\label{sec:density}

We model a starless core as having a spherically-symmetric density distribution which is a continuous function of radius $r$, defined by a central density $\rho_{c}$ and a characteristic size scale $R$,
\begin{equation}
\rho(\rho_{c},R,r)=\rho_{c}\times f(R,r)
\label{eq:density}
\end{equation}
where $f(R,r)$ is a monotonically decreasing function which obeys the limits $f(R,r\to 0)\to1$ and $f(R,r\to \infty)\to 0$.

We assume that the core is bounded by external pressure at a density of $\rho_{e}$ at a radius $r_{e}$ (where $r_{e}$ is a function of $\rho_{c}$, and $R$).  Material at radii $r<r_{e}$ is considered to belong to the core, and obeys the density relation given in equation~\ref{eq:density}.  Material at radii $r>r_{e}$ is considered to belong to the surrounding cloud, and has a constant density $\rho_{e}$.  Thus, the density profile of the system is given by
\begin{equation}
\rho(r)=\begin{cases}
\rho(\rho_{c},R,r) & 0<r\leq r_{e} \\
\rho_{e} & r\geq r_{e}.
\end{cases}
\end{equation}
Note that while we have assumed that this function is continuous across the boundary at $r_{e}$ (i.e. $\rho(\rho_{c},R,r_{e})=\rho_{e}$), there is no requirement in the model for this to be the case.

The mass of the core -- i.e. the mass enclosed in the radius $r_{e}$ -- is given by
\begin{equation}
M(r_{e})=4\uppi\int_{0}^{r_{e}}{\rm d}r\,r^{2}\rho(\rho_{c},R,r).
\label{eq:mass}
\end{equation}
We assume that the mass of the core remains constant throughout its evolution.

For a given characteristic radius $R$, and assuming a fixed enclosed mass $M$ and bounding density $\rho_{e}$, equation~\ref{eq:mass} can be solved for central density $\rho_{c}$, and hence for bounding radius $r_{e}$.  We thus construct the functions
\begin{equation}
\rho_{c}=\rho_{c}(M,\rho_{e},R)
\end{equation}
and
\begin{equation}
r_{e}=r_{e}(M,\rho_{e},R)
\end{equation}
for the behaviour of central density $\rho_{c}$ and bounding radius $r_{e}$ as a function of characteristic radius $R$, respectively.

In this work, we consider two cases in which equation~\ref{eq:mass} has an analytic solution.  However, this method could be generalised to non-analytic core geometries.

\subsection{Terms in the virial equation}

In Section~\ref{sec:density}, we showed that given our specifications for a core's density profile, for a given mass and bounding density, the bounding radius and central density of the core can be expressed as functions of characteristic radius only.  We consider the virial stability of the core as a function of characteristic radius, in order to determine whether the core is likely to contract or expand.  Contraction is defined as a decrease in the \emph{characteristic} radius $R$, while expansion is defined as an increase in characteristic radius $R$.  Note that this does not necessarily equate to identical behaviour in the bounding radius $r_{e}$.

We assess the stability of the core against contraction or expansion by estimating the magnitude of each of the terms in the virial equation.  We consider the virial equation in the form
\begin{equation}
\frac{1}{2}\ddot{\mathcal{I}}=2\Omega_{\textsc{k}}+\Omega_{\textsc{g}}+\Omega_{\textsc{m}}+\Omega_{\textsc{p}}
\label{eq:full_virial_theorem}
\end{equation}
where $\ddot{\mathcal{I}}$ is the second derivative of the moment of inertia $\mathcal{I}$, $\Omega_{\textsc{k}}$ is the internal energy, $\Omega_{\textsc{g}}$ is the gravitational potential energy, $\Omega_{\textsc{m}}$ is the magnetic energy, and $\Omega_{\textsc{p}}$ is the energy due to external pressure acting on the core.  If $\ddot{\mathcal{I}}<0$, a core's net energy is negative, and hence the core is virially bound.  Conversely, a core with $\ddot{\mathcal{I}}>0$ will be virially unbound, and the virially stable mass of a core is the mass at which $\ddot{\mathcal{I}}=0$.

We consider the virial stability of a core using the ratio
\begin{equation}
{\rm Virial\,\,Ratio} = -\frac{\Omega_{\textsc{g}}+\Omega_{\textsc{p}}}{2\Omega_{\textsc{k}}+\Omega_{\textsc{m}}},
\label{eq:virial_ratio}
\end{equation}
where a ratio value $>1$ indicates that the core is virially bound, a value $<1$ indicates that the core is virially unbound, and a value of 1 indicates that the core is virially stable.

\subsubsection{Gravitational potential energy}

The gravitational potential energy of the core is given by
\begin{equation}
\Omega_{\textsc{g}}(\rho_{c},R,r_{e})=-4\uppi G\int_{0}^{r_{e}}{\rm d}r\,r\,\rho(\rho_{c},R,r)M(\rho_{c},R,r), 
\label{eq:gpe}
\end{equation}
where $M(\rho_{c},R,r)$ is given by
\begin{equation}
M(\rho_{c},R,r)=4\uppi\int_{0}^{r}{\rm d}r^{\prime}\,r^{\prime2}\rho(\rho_{c},R,r^{\prime}).
\label{eq:mass_general}
\end{equation}

\subsubsection{External pressure energy}

The external pressure term in the virial equation, $\Omega_{\textsc{p}}$, is given by
\begin{equation}
\Omega_{\textsc{p}}=-3P_{\textsc{ext}}V=-4\uppi P_{\textsc{ext}}r_{e}^{3}
\label{eq:pressure_term}
\end{equation}
for a core of volume $V$ being acted on by an external pressure $P_{\textsc{ext}}$.  $P_{\textsc{ext}}$ can be estimated from the ideal gas law:
\begin{equation}
P_{\textsc{ext}}\approx\rho_{e}\sigma_{\textsc{ext}}^{2},
\label{eq:ideal_gas}
\end{equation}
where $\sigma_{\textsc{ext}}$ is the mean line-of-sight (one-dimensional) gas velocity dispersion in the material surrounding the core.

\begin{figure}
\centering
\includegraphics[width=0.47\textwidth]{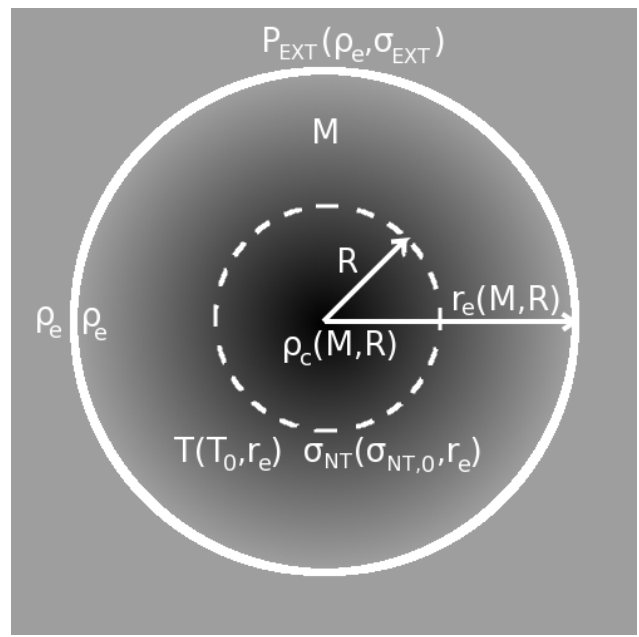}
\caption{Key parameters of the model; greyscale indicates local volume density, with darker shading indicating increased density.  The material interior to the white circle is considered to be part of the starless core.}
\label{fig:model}
\end{figure}

We assess the balance of confining forces -- whether the core is confined by external pressure, or by self-gravity -- with the ratio
\begin{equation}
{\rm Confinement\,\,Ratio}=\frac{\Omega_{\textsc{g}}}{\Omega_{\textsc{p}}},
\label{eq:omegag/omegap}
\end{equation}
where a ratio $>1$ indicates that the gravitational potential energy of the core is greater than the external pressure energy (i.e. the core is gravitationally confined), and a ratio $<1$ indicates that the external pressure energy is greater than the gravitational potential energy (i.e. the core is pressure-confined).

\subsubsection{Internal kinetic energy}

The internal energy term in the virial equation is given by
\begin{equation}
 \Omega_{\textsc{k}}=\frac{3}{2}M\sigma^{2}
\label{eq:kinetic_energy}
\end{equation}
where $\sigma$ is the line-of-sight (one-dimensional) velocity dispersion for the mean gas particle in the core.

\subsection{Adiabatic and isothermal variation of internal velocity dispersion}

The velocity dispersion $\sigma$ obeys the relation
\begin{equation}
\sigma^{2}(T,r_{e})=\sigma_{\textsc{t}}(T,r_{e})^{2}+\sigma_{\textsc{nt}}(r_{e})^{2},
\label{eq:sigma_ev}
\end{equation}
where $\sigma_{\textsc{t}}$ is the one-dimensional line-of-sight thermal gas velocity dispersion,
\begin{equation}
\sigma_{\textsc{t}}=\sqrt{\frac{k_{\textsc{b}}T}{\mu m_{\textsc{h}}}},
\end{equation}
and $\sigma_{\textsc{nt}}$ is the non-thermal gas velocity dispersion.

We model the evolution of the thermal velocity dispersion with core radius in the limits of isothermal and adiabatic compression of the core material.  In the isothermal case, the thermal velocity dispersion is given by
\begin{equation}
\sigma_{\textsc{t}}(T)=\sigma_{\textsc{t}}(T_{0}),
\label{eq:sigmat_iso_ev}
\end{equation}
where $T_{0}$ is the initial temperature of the gas in the starless core.  (All variables subscripted with a `0' refer to the initial value of that quantity.)  We assume $T_{0}=7$\,K, as being representative of the central temperature of a prestellar core (see, e.g. \citealt{stamatellos2007}).

In the adiabatic case, we assume that the gas obeys the adiabatic equation of state,
\begin{equation}
PV^{\gamma}=P_{0}V_{0}^{\gamma},
\label{eq:adiabatic}
\end{equation}
where $\gamma=7/5$, assuming the gas is diatomic.  If the gas is ideal then $PV\propto T$, and the equation of state becomes
\begin{equation}
T=T_{0}\left(\frac{V_{0}}{V}\right)^{\gamma-1}=T_{0}\left(\frac{V_{0}}{V}\right)^{\frac{2}{5}}=T_{0}\left(\frac{r_{e}}{r_{e,0}}\right)^{-\frac{6}{5}}. 
\label{eq:sigmat_adi_ev}
\end{equation}

We assume that turbulence dissipates as a core contracts, and that the non-thermal component of the linewidth decreases in a manner which obeys the \citet{solomon1987} relation between the size and non-thermal internal linewidth of a starless core,
\begin{equation}
\sigma_{\textsc{nt}}\propto r_{e}^{0.5}.
\label{eq:larsonlike}
\end{equation}
We discuss the validity of this assumption in Section~\ref{sec:nonthermal}.  We further assume that there is no mechanism by which a core can increase its non-thermal velocity dispersion, as there is no mechanism for injecting turbulence into the system.  Thus, we parameterise the non-thermal linewidth as
\begin{equation}
\sigma_{\textsc{nt}}=\begin{cases}
\sigma_{\textsc{nt},0}\left(\frac{r_{e}}{r_{e,0}}\right)^{0.5} & r_{e}\leq r_{e,0} \\
\sigma_{\textsc{nt},0} & r_{e}>r_{e,0}.
\end{cases}
\label{eq:sigmant_ev}
\end{equation}
This is the only non-reversible parameter in our model.  Note that this simple functional form assumes that $r_{e}$ is a monotonic function of $R$, i.e. that a core which has decreased in size and hence dissipated its turbulence will not subsequently increase in size again.  This condition is true for the density profiles which we consider in this paper.

\subsection{Model parameters}

Figure~\ref{fig:model} shows the key parameters of the model, for a generalised starless core density distribution with characteristic radius $R$.  Our model requires the specification of the following measurable initial conditions: a total core mass $M$ (fixed throughout), the density of the surrounding medium $\rho_{e}$ (fixed throughout), an initial characteristic radius $R_{0}$, an initial temperature $T_{0}$, an initial internal gas velocity dispersion $\sigma_{0}$, and an external gas velocity dispersion $\sigma_{\textsc{ext}}$ (fixed throughout).

For a given set of initial (measurable) conditions, and having chosen an appropriate core density profile, we can create adiabatic and isothermal evolutionary tracks as a function of core characteristic radius $R$ only.

\subsection{Model assumptions}

Throughout this analysis, we assume that the enclosed core mass $M$ is fixed (i.e. that there is negligible accretion of mass onto the core), that the core density profile obeys the distribution given in equation~\ref{eq:density} at all radii less than the truncation radius $r_{e}$, that the core has a uniform temperature at all radii less than the truncation radius (although this temperature varies as a function of characteristic size in the case of adiabatic evolution), that the core is bounded by a constant external pressure $P_{\textsc{ext}}$ at a constant density $\rho_{e}$, and that no external disruption occurs.

\section{Analytic density profiles}

\subsection{Plummer-like density profile}

Starless cores are frequently modelled as having Plummer-like density distributions (e.g. \citealt{whitworth2001}).  This model produces a density profile consistent with observations showing that starless cores typically have a flat central plateau and a power-law decrease in density at large radii (e.g. \citealt{alves2001}).  It also produces a density profile similar to that of a Bonnor-Ebert distribution (\citealt{ebert1955}; \citealt{bonnor1956}).  There are a family of Plummer-like distributions, characterised by their power-law behaviour at large radii \citep{plummer1911}, some of which have fully analytic solutions.  We consider one of these analytic solutions here.

If the starless core obeys a Plummer-like density distribution \citep{plummer1911}, then $f(R,r)$ is given by
\begin{equation}
f(R,r)= \left(\frac{R}{\sqrt{r^{2}+R^{2}}}\right)^{\eta},
\end{equation}
where $R$ defines the radius of the flat central plateau of the distribution and $\eta$ defines the power-law slope of the distribution at large radii.  For a true Plummer distribution, $\eta=5$ \citep{plummer1911}.  Throughout the remainder of this work we will assume $\eta=4$ \citep{whitworth2001}.  Thus, equation~\ref{eq:density} becomes
\begin{equation}
\rho(\rho_{c},R,r)=\rho_{c}\times \left(\frac{R}{\sqrt{r^{2}+R^{2}}}\right)^{4}.
\label{eq:density_p}
\end{equation}

The core is truncated at a radius
\begin{equation}
r_{e}=R\sqrt{\left(\frac{\rho_{c}}{\rho_{e}}\right)^{\frac{1}{2}}-1},
\label{eq:re_p}
\end{equation}
at which the density drops to the bounding value, $\rho_{e}$.

The mass of a truncated Plummer-like distribution when $\eta=4$ is given by
\begin{equation}
M(r_{e})=2\uppi\rho_{c}R^{3}\left[\arctan\left(\frac{r_{e}}{R}\right)-\frac{r_{e}R}{r_{e}^{2}+R^{2}}\right].
\label{eq:mass_trunc_p}
\end{equation}
For a given mass $M$, bounding density $\rho_{e}$, and flat radius $R$, this equation can, when coupled with equation~\ref{eq:re_p}, be solved numerically for core central density $\rho_{c}$.

The gravitational potential energy of this distribution is given by
\begin{multline}
\Omega_{\textsc{g}}=-\uppi^{2}G\rho_{c}^{2}R^{7}\left[\frac{2r_{e}R}{(r_{e}^{2}+R^{2})^{2}}\right. \\ +\left.\frac{1}{R^{2}}\left(\arctan\left(\frac{r_{e}}{R}\right)+\frac{r_{e}R}{r_{e}^{2}+R^{2}}\right)-\frac{4\arctan\left(\frac{r_{e}}{R}\right)}{r_{e}^{2}+R^{2}}\right].
\label{eq:gpe_trunc_p}
\end{multline}
These results are derived in Appendix A.

As $r_{e}/R\to\infty$ (which in this model occurs when both $r_{e}$ and $R$ become small, as discussed below), equation~\ref{eq:gpe_trunc_p} tends to the gravitational potential energy of an infinite $\eta=4$ Plummer-like distribution,
\begin{equation}
  \Omega_{\textsc{g},{\rm inf}}=-\frac{1}{2\uppi}\frac{GM_{\rm inf}^{2}}{R},
\label{eq:gpe_p}
\end{equation}
where $M_{\rm inf}$ is the mass of an infinite $\eta=4$ Plummer-like distribution,
\begin{equation}
M_{\rm inf} = \uppi^{2}\rho_{c}R^{3}.
\end{equation}

\subsection{Gaussian density profile}

Starless cores have previously been modelled as having Gaussian density distributions (e.g. \citealt{wardthompson1994}).  This model has the advantage of being particularly analytically tractable.

If the starless core obeys a Gaussian density profile, then $f(R,r)$ is given by
\begin{equation}
f(R,r)=e^{-\frac{1}{2}\left(\frac{r}{R}\right)^{2}},
\end{equation}
where $R$ is the characteristic radius of the Gaussian distribution, and equation~\ref{eq:density} becomes
\begin{equation}
\rho(\rho_{c},R,r)=\rho_{c}\times e^{-\frac{1}{2}\left(\frac{r}{R}\right)^{2}}.
\label{eq:density_g}
\end{equation}

The core is truncated at a radius
\begin{equation}
r_{e}=R\sqrt{2{\rm ln}\left(\frac{\rho_{c}}{\rho_{e}}\right)},
\label{eq:re_g}
\end{equation}
at which the density drops to the bounding value, $\rho_{e}$.

The mass enclosed by a truncated Gaussian distribution is given by
\begin{equation}
M(r_{e})=4\uppi\rho_{c}\left[R^{3}\sqrt{\frac{\uppi}{2}}\,{\rm erf}\left(\frac{r_{e}}{R\sqrt{2}}\right)-R^{2}r_{e}e^{\nicefrac{-r_{e}^{2}}{2R^{2}}}\right],
\label{eq:mass_trunc_g}
\end{equation}
where `$\erf$' is the error function.  For a given mass $M$, bounding density $\rho_{e}$, and characteristic radius $R$, this equation can, when coupled with equation~\ref{eq:re_g}, be solved numerically for core central density $\rho_{c}$.

The gravitational potential energy of a truncated Gaussian distribution is given by
\begin{multline}
\Omega_{\textsc{g}}=-16\uppi^{2}G\rho_{c}^{2}R^{5}\left[\frac{\sqrt{\uppi}}{4}\erf\left(\frac{r_{e}}{R}\right)\right. \\ -\left.\sqrt{\frac{\uppi}{2}}e^{-\frac{1}{2}\left(\frac{r_{e}}{R}\right)^{2}}\erf\left(\frac{r_{e}}{R\sqrt{2}}\right)+\frac{1}{2}\frac{r_{e}}{R}e^{-\left(\frac{r_{e}}{R}\right)^{2}}\right].
\label{eq:gpe_trunc_g}
\end{multline}
These results are derived in Appendix B.

As $r_{e}/R\to\infty$, equation~\ref{eq:gpe_trunc_g} tends to the gravitational potential energy of an infinite Gaussian distribution,
\begin{equation}
  \Omega_{\textsc{g},{\rm inf}}=-\frac{1}{2\sqrt{\uppi}}\frac{GM_{\rm inf}^{2}}{R}
\label{eq:gpe_g}
\end{equation}
(see \citealt{pattle2015} for a derivation of this result), where $M_{\rm inf}$ is the mass of an infinite Gaussian distribution,
\begin{equation}
M_{\rm inf} = 2\sqrt{2}\uppi^{\nicefrac{3}{2}}\rho_{c}R^{3}.
\label{eq:mass_p}
\end{equation}

\section{Core stability as a function of characteristic radius}

We initially solve equations~\ref{eq:virial_ratio} and \ref{eq:omegag/omegap} for non-magnetic starless cores with with masses 0.1\,M\sun, 0.25\,M\sun\ and 0.5\,M\sun, external pressure $P_{\textsc{ext}}/k_{\textsc{b}}=1.5\times10^{7}$\,K\,cm$^{-3}$, and, at an initial characteristic radius of 0.005\,pc, a core temperature of 7\,K and a non-thermal internal velocity dispersion of 250\,m\,s$^{-1}$.  These values are chosen to be representative of starless cores in the Ophiuchus molecular cloud (see Table~\ref{tab:ics}, below).  We assume that the cores are confined by material at a density $\rho_{e}=10^{5}$ H$_{2}$\,molecules\,cm$^{-3}$, and that the mean molecular weight is 2.86$\times m_{\textsc{h}}$ (assuming that the core is 70\% molecular hydrogen by mass; \citealt{kirk2013}). These values were chosen in order to illustrate the range of behaviours predicted by our model for a typical low-mass starless core.

\begin{figure*}
\centering
\includegraphics[width=\textwidth]{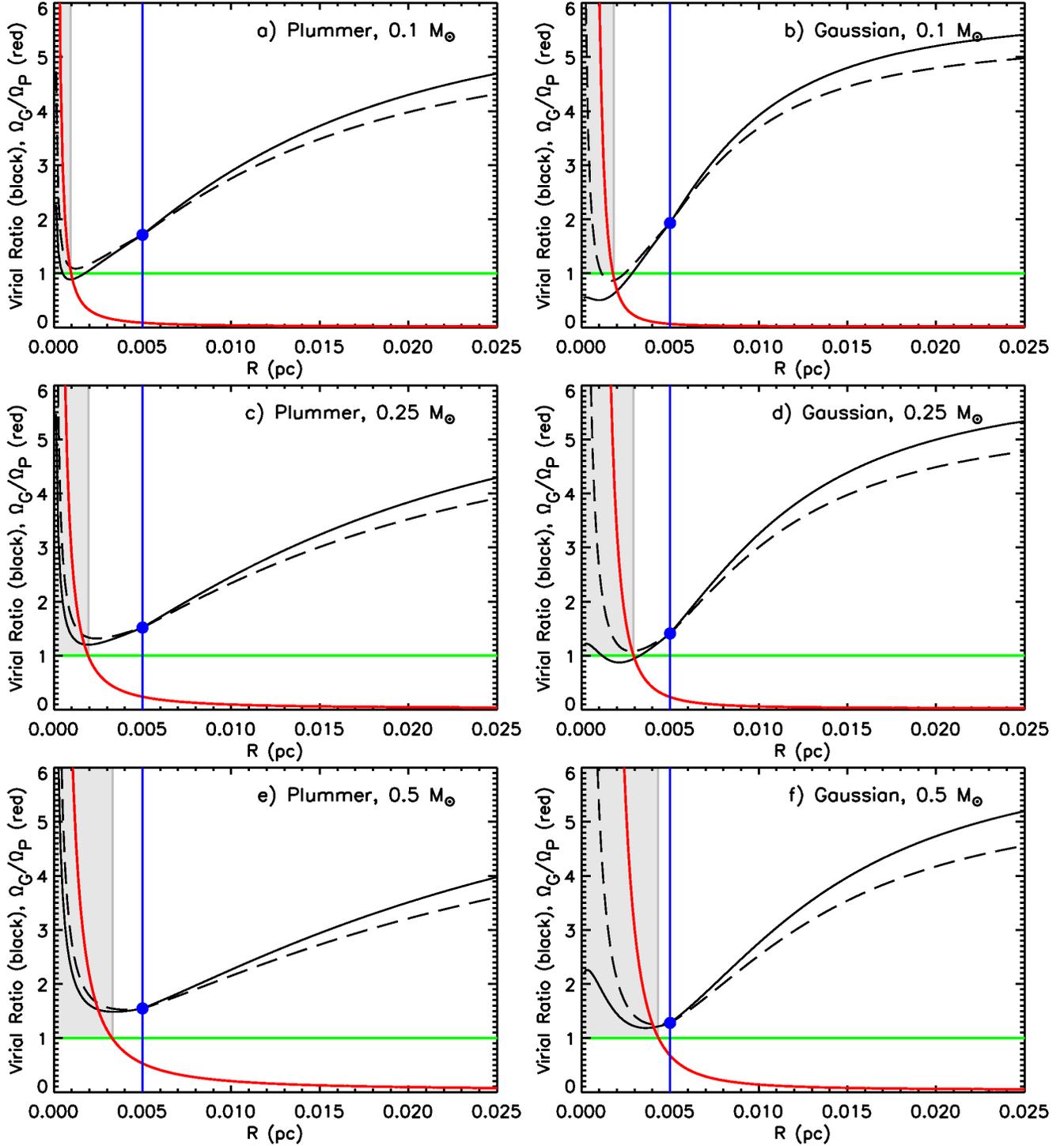}
\caption{Stability as a function of characteristic radius for hypothetical starless cores of mass 0.1\,M$_{\odot}$, 0.25\,M$_{\odot}$ and 0.5\,M$_{\odot}$, (see text for details of core properties).  The left-hand column assumes a Plummer-like density profile; the right-hand column assumes a Gaussian density profile.  Solid black line shows adiabatic virial ratio; dashed black line shows isothermal virial ratio.  Red line shows ratio of gravitational potential energy to external pressure energy.  Blue dot shows initial virial ratio (vertical blue line is to guide the eye to the initial value of the confinement ratio).  Horizontal green line shows line of virial stability.  Grey shaded region indicates parameter space in which the core would be considered to be prestellar.}
\label{fig:hypothetical}
\end{figure*}

Figure~\ref{fig:hypothetical} shows the virial ratio, $-(\Omega_{\textsc{g}}+\Omega_{\textsc{p}})/2\Omega_{\textsc{k}}$, in black and the confinement ratio, $\Omega_{\textsc{g}}/\Omega_{\textsc{p}}$, in red, both plotted as a function of core characteristic radius $R$.  The left-hand column of Figure~\ref{fig:hypothetical} shows the result of assuming a Plummer-like density distribution, while the right-hand column shows the result of a Gaussian density distribution.  The solid black line shows the virial ratio in the adiabatic case, while the dashed black line shows the virial ratio in the isothermal case.  The blue point marks the initial virial ratio of the core, while the blue line extending from that point guides the eye to the initial value of $\Omega_{\textsc{g}}/\Omega_{\textsc{p}}$ on the red curve.  The green line marks where the virial and confinement ratios equal unity.  While the virial ratio is above the green line the core is virially bound, and while it is below, the core is unbound.  Similarly, while the confinement ratio is above the green line, the core is gravitationally-dominated; while it is below the green line it is pressure-dominated.  The region shaded in grey is `prestellar': where $-(\Omega_{\textsc{g}}+\Omega_{\textsc{p}})/2\Omega_{\textsc{k}}<1$ and $\Omega_{\textsc{g}}/\Omega_{\textsc{p}}>1$.  A core whose virial ratio (the black line) falls in this region will be both virally bound and confined by gravity.

As can be seen in Figure~\ref{fig:hypothetical}, the results of the two density profiles are qualitatively very similar.  In both cases, our model for the virial parameter as a function of $R$ shows two regimes.  The first is a gravitationally-dominated regime at small $R$.  If this regime is not virially bound over all of its range, it generally becomes so as $R$ approaches zero (with the exception of adiabatic collapse in some Gaussian models, discussed below).  The second is a pressure-dominated regime at large $R$ which will be virially bound over some or all of its range, and will become increasingly virially bound as $R$ increases.

\begin{figure*}
\centering
\includegraphics[width=0.9\textwidth]{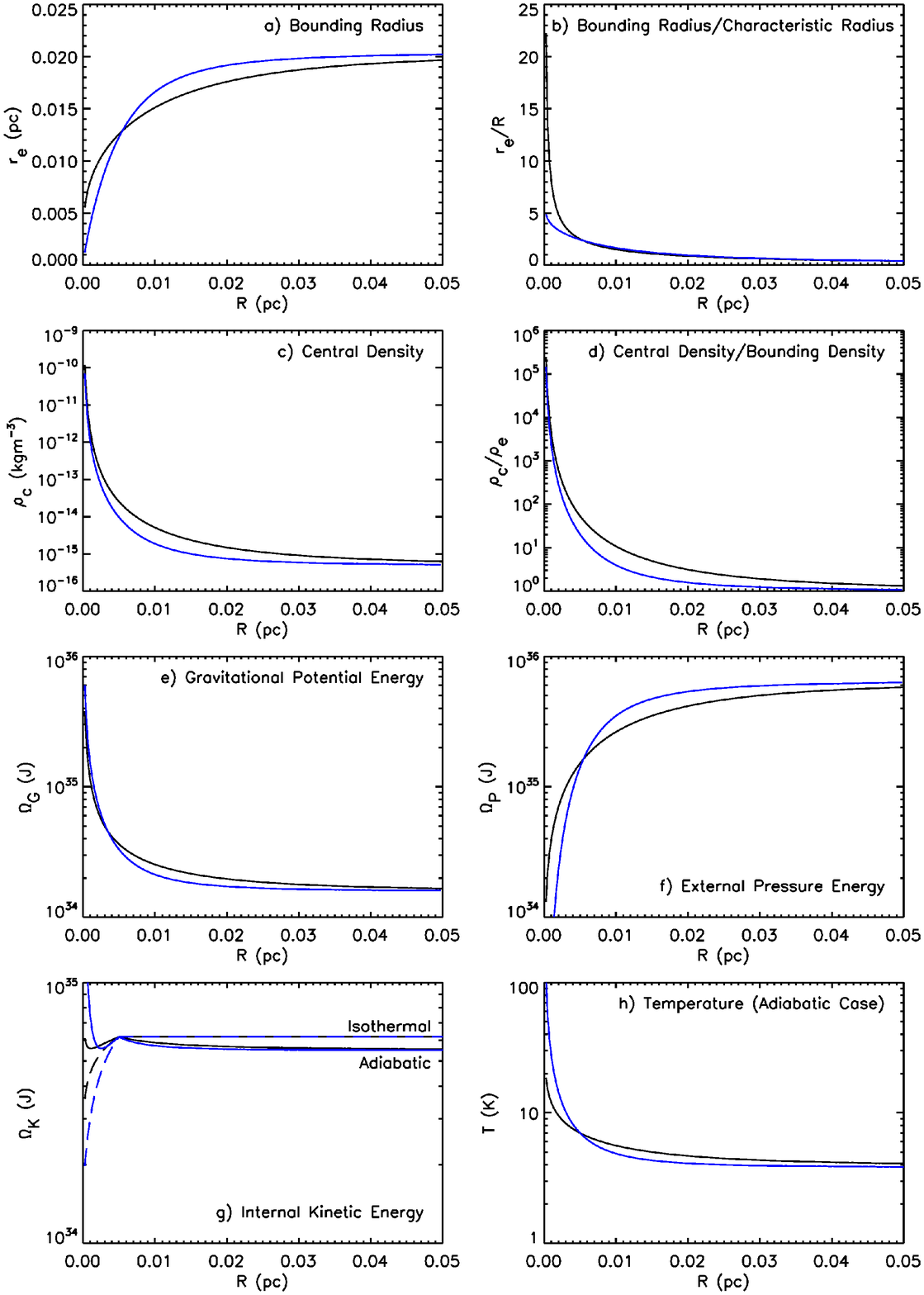}
\caption{Various model terms as a function of core characteristic radius $R$ for a core of mass 0.25\,M\sun\ (see text for details of core properties), assuming a Plummer-like density profile (black), and a Gaussian density profile (blue): a) Bounding radius, $r_{e}$; b) Ratio of bounding radius to characteristic radius, $r_{e}/R$; c) Central density, $\rho_{c}$; d) Ratio of central and bounding densities, $\rho_{c}/\rho_{e}$; e) Gravitational potential energy, $\Omega_{\textsc{g}}$; f) External pressure energy, $\Omega_{\textsc{p}}$; g) Internal kinetic energy (solid line -- adiabatic, dashed line -- isothermal), $\Omega_{\textsc{k}}$; h) Temperature in the adiabatic case, $T$.}
\label{fig:example}
\end{figure*}

We show the behaviour of various terms in our model as a function of characteristic radius in Figure~\ref{fig:example} for the core of mass 0.25\,M$_{\odot}$ described above.  We use this figure to explain the behaviour of the virial and confinement ratios seen in Figure~\ref{fig:hypothetical}.

Figures~\ref{fig:example}a and \ref{fig:example}b show the behaviour of the bounding radius $r_{e}$ as a function of characteristic radius $R$.  When $R$ is small, the bounding radius is much greater than the characteristic radius, and so the gravitational potential energy of the core tends to the value it would take if the core were infinite in extent.  As the core's characteristic and bounding radii increase, in order to conserve mass, the central density decreases (see Figure~\ref{fig:example}b).  Similarly, the density contrast between centre and edge decreases (see Figure~\ref{fig:example}d), and the ratio of the bounding radius to the characteristic radius becomes small (see equations~\ref{eq:re_p} and \ref{eq:re_g}).  At large values of $R$, the bounding radius $r_{e}$ is much smaller than the characteristic radius and the density contrast from the centre to the edge of the core becomes small, and so the gravitational potential energy tends toward the value it would take if the core were a uniform sphere of radius $r_{e}$.  The behaviour of gravitational potential energy as a function of $R$ is shown in Figure~\ref{fig:example}e.

At the smallest $R$, gravity dominates over external pressure, as $\Omega_{\textsc{g}}\propto R^{-1}$, and $\Omega_{\textsc{p}}\propto r_{e}^{3}$.  The variation of external pressure energy with core characteristic radius $R$ is shown in Figure~\ref{fig:example}f.  At small $R$ the total energy due to external pressure is small because the source is small.  At intermediate $R$, $\Omega_{\textsc{p}}$ increases, while $\Omega_{\textsc{g}}$ falls off as shown in Figure~\ref{fig:example}e, and so the core becomes pressure-dominated.  $\Omega_{\textsc{p}}$ continues to dominate at large $R$, as the gravitational potential energy approaches that of a uniform sphere, and as $\Omega_{\textsc{p}}$ increases with $r_{e}$.

The internal kinetic energy of the core stays approximately constant over a wide range of radii (Figure~\ref{fig:example}g).  The differences between isothermal and adiabatic behaviour are significant at small radii only, discussed below.

\subsection{Behaviour of model at small characteristic radii}

The behaviour of the Plummer-like and Gaussian core geometries diverge significantly only at small radii, as shown in Figure~\ref{fig:example}.  This is due to a core with a Gaussian density distribution being more centrally condensed than a core with a Plummer-like density distribution.  In both cases, when characteristic radius $R$ is small the bounding radius $r_{e}$ is also small, but is significantly larger than the characteristic radius.  At small radii the bounding radius of the Plummer-like distribution is larger than that of the Gaussian distribution, in order to conserve the mass enclosed $M$ while obeying a less centrally condensed density distribution.  Thus, the Plummer-like distribution has a higher energy due to external pressure at small radii than the Gaussian distribution, as $\Omega_{\textsc{p}}\propto r_{e}^{3}$ (see Figure~\ref{fig:example}f).

For both density distributions the gravitational potential energy increases as $R$ becomes small, tending toward the value of $\Omega_{\textsc{g}}$ would take if the core were infinite in extent.  The Gaussian model tends toward a value of $\Omega_{\textsc{g}}$ which is $\sqrt{\uppi}$ greater than that of the Plummer-like model (compare equations~\ref{eq:gpe_p} and \ref{eq:gpe_g}, and see Figure~\ref{fig:example}e).  Hence, the Gaussian model becomes gravitationally dominated ($\Omega_{\textsc{g}}>\Omega_{\textsc{p}}$) at larger values of $R$ than the Plummer-like model (compare areas shaded grey in Figure~\ref{fig:hypothetical}).  However, the Plummer-like model is more virially bound than the Gaussian model at small-to-intermediate radii, due to the dependence of the external pressure energy of the core on bounding radius (see Figure~\ref{fig:hypothetical}).
 
At small radii the adiabatic and isothermal values of the internal kinetic energy diverge significantly, as shown in Figure~\ref{fig:example}g.  At small values of $r_{e}$, the contribution of the non-thermal kinetic energy becomes small (see equation~\ref{eq:sigmant_ev}), and the behaviour of the total internal kinetic energy is dominated by the behaviour of the thermal kinetic energy term.  Thus, in the isothermal case, at small radii the internal kinetic energy tends to a constant value,
\begin{equation}
\Omega_{\textsc{k,i}}\to \frac{3}{2}M\frac{k_{\textsc{b}}}{\mu m_{\textsc{h}}}T_{0}.
\label{eq:small_r_omegaki}
\end{equation}
$\Omega_{\textsc{k,i}}$ tends to this value when
\begin{equation}
\frac{r_{e}}{r_{e,0}}\ll\frac{k_{\textsc{b}}T_{0}}{\mu m_{\textsc{h}}\sigma_{\textsc{nt,0}}^{2}}.
\end{equation}
The core shown in Figure~\ref{fig:example} has a large initial non-thermal linewidth, 250 m\,s$^{-1}$, and so in this case, $\Omega_{\textsc{k,i}}$ tends to a constant value at radii smaller than those shown in Figure~\ref{fig:example}g.

The adiabatic internal kinetic energy will diverge at small radii, as
\begin{equation}
\Omega_{\textsc{k,a}}\to \frac{3}{2}M\frac{k_{\textsc{b}}}{\mu m_{\textsc{h}}}T_{0}\left(\frac{r_{e}}{r_{e,0}}\right)^{-\frac{6}{5}}.
\end{equation}
The adabatic internal kinetic energy will become larger faster as $R\to 0$ in the Gaussian case than the Plummer-like case, as $r_{e}$ becomes smaller faster with decreasing $R$ in the Gaussian case (see Figures~\ref{fig:example}a and \ref{fig:example}g).  Thus, in the Gaussian model, the virial ratio in the adiabatic case may tend to a value $< 1$ -- i.e. the core may be virially unbound -- as $R$ becomes small (see Figure~\ref{fig:hypothetical}b).

The increase in adiabatic internal kinetic energy at small radii is equivalent to an increase in core temperature, as the core heats as it collapses.  This increase is shown in Figure~\ref{fig:example}h, in which it can be seen that the Gaussian model results in higher core temperatures at small $R$ than the Plummer-like model.  The Gaussian model shows core temperatures $\sim 100\,$K at the smallest radii, while the Plummer-like model shows a more physically plausible increase in temperature, up to $\sim 20\,$K.

The physical relevance of quasistatic contraction of the core at the smallest radii is not certain.  Once a core is both virially bound and gravitationally dominated, the quasistatic model is unlikely to apply.  However, in this case, the core is still expected to undergo further collapse.  The simple model presented in this work is only justified over the range of characteristic radii which have been measured for starless cores, and may not be relevant at the smallest and largest radii.

\subsection{Behaviour of model at large characteristic radii}

Both the Plummer-like and the Gaussian core geometries tend to the same behaviour at large radii, as shown in Figure~\ref{fig:example}.

\begin{figure*}
\centering
\includegraphics[width=\textwidth]{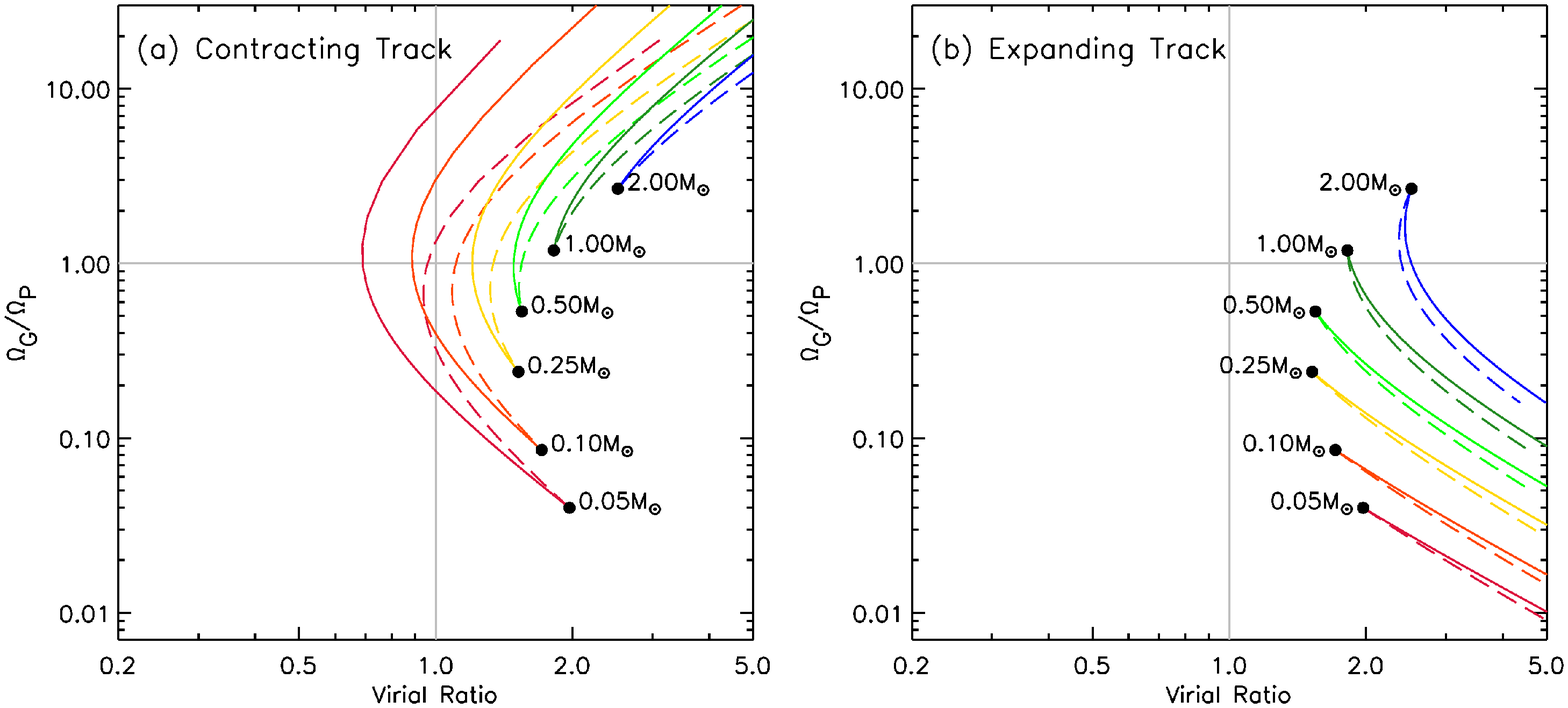}
\caption{A family of loci of Equations~\ref{eq:density}--\ref{eq:sigmant_ev} in the virial plane, assuming a Plummer-like core geometry, showing (a) the contracting track and (b) the expanding track.  Solid lines show adiabatic loci; dashed lines show isothermal loci.  Cores have $M=0.05-2.0$\,M\sun, external pressure $P_{\textsc{ext}}/k_{\textsc{b}}=1.5\times10^{7}$\,K\,cm$^{-2}$, an initial temperature of 7\,K, an initial non-thermal linewidth of 250\,m\,s$^{-1}$, and an initial characteristic radius of 0.005\,pc, and are confined by material of density $\rho_{e}=10^{5}$ H$_{2}$\,molecules\,cm$^{-3}$.  For clarity, each mass track has a different colour.}
\label{fig:evolution_example_1}
\includegraphics[width=\textwidth]{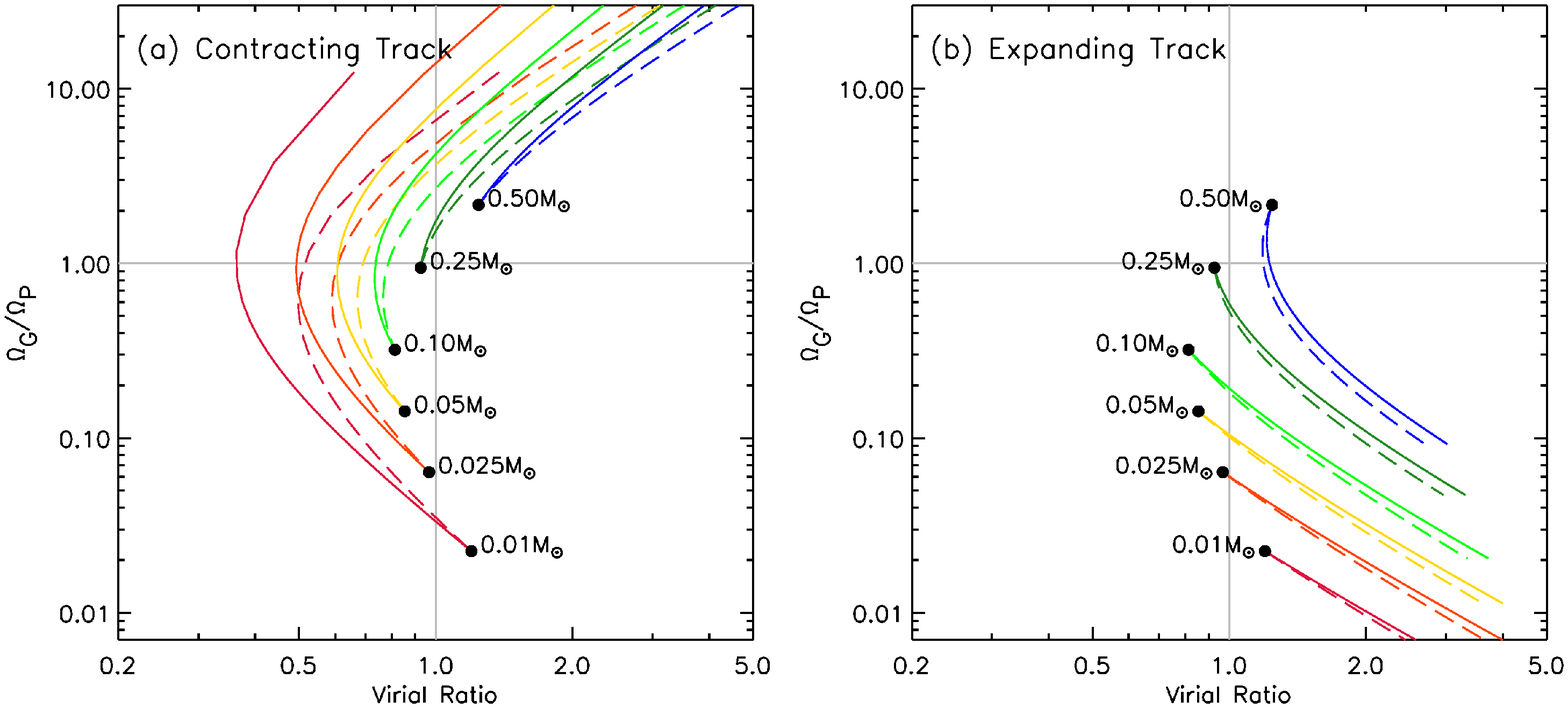}
\caption{A family of loci of Equations~\ref{eq:density}--\ref{eq:sigmant_ev} in the virial plane, assuming a Plummer-like core geometry, showing (a) the contracting track and (b) the expanding track.  Solid lines show adiabatic loci; dashed lines show isothermal loci.  Cores have $M=0.001-0.5$\,M\sun, external pressure $P_{\textsc{ext}}/k_{\textsc{b}}=1.5\times10^{7}$\,K\,cm$^{-2}$, an initial temperature of 7\,K, an initial non-thermal linewidth of 280\,m\,s$^{-1}$, and an initial characteristic radius of 0.002\,pc, and are confined by material of density $\rho_{e}=10^{5}$ H$_{2}$\,molecules\,cm$^{-3}$.  For clarity, each mass track has a different colour.}
\label{fig:evolution_example_2}
\end{figure*}

Figure~\ref{fig:example}a shows that as $R$ becomes large, the rate of increase of $r_{e}$ with $R$ becomes small.  This is because as the centre-to-edge density contrast approaches unity, and the core density distribution approaches that a uniform sphere, the density interior to the core becomes insensitive to changes in $R$, and $r_{e}$ tends toward the value it would take if the core were a uniform sphere of density $\rho_{e}$ and mass $M$, i.e. $r_{e}\to (3M/4\uppi\rho_{e})^{\nicefrac{1}{3}}$.  The gravitational potential energy of the core tends toward
\begin{equation}
\Omega_{\textsc{g}}\to-\frac{3}{5}\frac{GM^{2}}{r_{e}}=-\frac{3}{5}\left(\frac{4\uppi}{3}\right)^{\frac{1}{3}}G\rho_{e}^{\frac{1}{3}}M^{\frac{5}{3}},
\end{equation}
while the external pressure energy tends toward
\begin{equation}
\Omega_{\textsc{p}}=-4\uppi P_{\textsc{ext}}r_{e}^{3}\to-\frac{3MP_{\textsc{ext}}}{\rho_{e}}.
\end{equation}
Thus, the confinement ratio $\Omega_{\textsc{g}}/\Omega_{\textsc{p}}$ tends toward
\begin{equation}
\frac{\Omega_{\textsc{g}}}{\Omega_{\textsc{p}}}\to \frac{1}{5}\left(\frac{4\uppi}{3}\right)^{\frac{1}{3}}GM^{\frac{2}{3}}\rho_{e}^{\frac{4}{3}}P_{\textsc{ext}}^{-1}.
\end{equation}
Figure~\ref{fig:example}g shows the behaviour of the internal kinetic energy of the core with $R$.  In the isothermal case, as the core expands, the internal kinetic energy maintains its initial value,
\begin{equation}
\Omega_{\textsc{k,i}}=\frac{3}{2}\left[\frac{k_{\textsc{b}}T_{0}}{\mu m_{\textsc{h}}}+\sigma_{\textsc{nt},0}^{2}\right],
\end{equation}
However, in the adiabatic case, the internal kinetic energy, $\Omega_{\textsc{k,a}}$, tends toward a smaller value,
\begin{equation}
\Omega_{\textsc{k,a}}\to\frac{3}{2}\left[\frac{k_{\textsc{b}}T_{0}}{\mu m_{\textsc{h}}r_{e,0}^{-\frac{6}{5}}}\left(\frac{3M}{4\uppi\rho_{e}}\right)^{-\frac{2}{5}}+\sigma_{\textsc{nt},0}^{2}\right].
\end{equation}
Thus, at large $R$, in the isothermal case the virial ratio tends toward the value
\begin{multline}
-\frac{\Omega_{\textsc{g}}+\Omega_{\textsc{p}}}{2\Omega_{\textsc{k,i}}}\to\frac{2}{3} \left[\frac{3}{5}\left(\frac{4\uppi}{3}\right)^{\frac{1}{3}}G\rho_{e}^{\frac{1}{3}}M^{\frac{5}{3}}+\frac{3MP_{\textsc{ext}}}{\rho_{e}}\right]    \\
\times\left[\frac{k_{\textsc{b}}T_{0}}{\mu m_{\textsc{h}}}+\sigma_{\textsc{nt},0}^{2}\right]^{-1},
\end{multline}
while in the adiabatic case the virial ratio tends toward the slightly larger value
\begin{multline}
-\frac{\Omega_{\textsc{g}}+\Omega_{\textsc{p}}}{2\Omega_{\textsc{k,a}}}\to\frac{2}{3} \left[\frac{3}{5}\left(\frac{4\uppi}{3}\right)^{\frac{1}{3}}G\rho_{e}^{\frac{1}{3}}M^{\frac{5}{3}}+\frac{3MP_{\textsc{ext}}}{\rho_{e}}\right]    \\
\times\left[\frac{k_{\textsc{b}}T_{0}}{\mu m_{\textsc{h}}r_{e,0}^{-\frac{6}{5}}}\left(\frac{3M}{4\uppi\rho_{e}}\right)^{-\frac{2}{5}}+\sigma_{\textsc{nt},0}^{2}\right]^{-1}.
\end{multline}

As $\rho_{c}\to\rho_{e}$, the core tends toward the behaviour of a uniform sphere, while becoming effectively indistinguishable from the medium in which it is embedded.  However, as discussed below, this parameter space is not physically accessible to the realistic starless cores which we consider. 

\subsection{Core evolution}
\label{sec:core_evolution}

When considering the evolution of the cores in our sample, we presume that any virially bound and gravitationally-dominated core ($-[\Omega_{\textsc{g}}+\Omega_{\textsc{p}}]/2\Omega_{\textsc{k}}>1$ and $\Omega_{\textsc{g}}/\Omega_{\textsc{p}}>1$) is prestellar and collapsing under gravity, and will evolve away from virial equilibrium to become more gravitationally bound -- i.e., we expect a core which occupies the grey-shaded regions of Figure~\ref{fig:hypothetical} to evolve toward smaller radii in all cases.  It is unlikely to do so precisely along the evolutionary track given by our model, as the core's evolution will not continue to be quasistatic as it undergoes runaway collapse under gravity.

We assume that a virially-bound and pressure-dominated core will contract under external pressure until it either reaches virial equilibrium or becomes gravitationally unstable.  An effect of the functional form of $\Omega_{\textsc{p}}$ is to produce a local minimum in the virial ratio in the intermediate region between gravitationally-dominated and pressure-dominated behaviour $(\Omega_{\textsc{g}}\sim\Omega_{\textsc{p}})$ at small $R$.  This minimum can be seen for every core in Figure~\ref{fig:hypothetical}.  As a result of this minimum, not all contracting pressure-confined and virially-bound cores will become gravitationally-bound prestellar cores.  In Figure~\ref{fig:hypothetical}, while all the cores are initially virially-bound and pressure-confined, the 0.25\,M\sun\ Plummer-like core (Figure~\ref{fig:hypothetical}c) and both the Gaussian and Plummer-like 0.5\,M\sun\ starless cores (Figures~\ref{fig:hypothetical}e and \ref{fig:hypothetical}f) will evolve to become prestellar in both the adiabatic and the isothermal cases, as in each of these cases, the core becomes gravitationally-dominated ($\Omega_{\textsc{g}}>\Omega_{\textsc{p}}$) while virially bound, and does not subsequently become virialised.  The Gaussian 0.1\,M\sun\ core in Figure~\ref{fig:hypothetical}b becomes virialised while still pressure-dominated in both the adiabatic and isothermal cases.  The 0.1\,M\sun\ Plummer-like core (Figure~\ref{fig:hypothetical}a) and the 0.25\,M\sun\ Gaussian core (Figure~\ref{fig:hypothetical}d) become gravitationally-dominated while virially bound in the isothermal case, but in the adiabatic case become virialised while pressure-dominated.

We expect an initially virially unbound core to expand due to its internal pressure until it reaches a pressure-bound virial equilibrium and, once that equilibrium is reached, to remain in or near virial equilibrium thereafter.  Note that if the core has an initial characteristic radius less than that at which the minimum in virial ratio occurs, this increase in radius will initially cause the core to expand \emph{away} from virial equilibrium.

We do not expect a starless core which contracts to equilibrium with its surroundings to instantaneously cease its contraction (or a core which expands to equilibrium to cease its expansion).  A more realistic scenario is one in which the core passes virial equilibrium, until the increasing virial instability forces its contraction (or expansion) to halt, and then reverse.  One might expect these pressure-confined starless cores without a route to gravitational instability to oscillate slightly around virial equilibrium.  Figures~\ref{fig:hypothetical}a, \ref{fig:hypothetical}b and \ref{fig:hypothetical}d show examples of cases in which there are characteristic radii $R$ at which the virial ratio of the core is predicted to be equal to 1.  Examination of those points where the virial ratio is equal to 1 and the core is pressure-dominated ($\Omega_{\textsc{g}}<\Omega_{\textsc{p}}$) shows that small perturbations in $R$ will have a tendency to force the core back towards virial equilibrium.  \citet{keto2006} suggested, and modelled, oscillating pressure-confined starless cores as an explanation for starless cores observed to show red-asymmetric line profiles, or reversals in line-profile asymmetry.

\subsection{Choice of core geometry for further analysis}

Figures~\ref{fig:hypothetical} and \ref{fig:example} show that the Plummer-like and Gaussian density profiles produce similar core behaviours.  For the remainder of this paper, we choose to model cores as obeying a Plummer-like density profile.  The Plummer-like density profile is more physically motivated than the Gaussian density profile, and produces a more physically realistic temperature range for the core.  The Plummer-like density profile can be seen as an analytically-soluble approximation to the Bonnor-Ebert density profile.

\section{Evolutionary tracks in the virial plane}

\citet{pattle2015} introduced the `virial plane' as a means of demonstrating the balance of forces in a starless core.  We use this diagram throughout the remainder of this paper, and so introduce it in some detail here.  The virial ratio is plotted as the abcissa, and the ratio of gravitational potential energy to external pressure energy (the confinement ratio) is plotted as the ordinate.  The virial ratio indicates the virial stability of the starless core, while the gravitational potential/external pressure energy ratio indicates the mode of core confinement.  Examples of the virial plane can be seen in Figures~\ref{fig:evolution_example_1} and \ref{fig:evolution_example_2}, discussed below.  Cores on the right-hand side of the virial plane (virial ratio $> 1$) are virially bound, while cores on the left-hand side of the virial plane (virial ratio $< 1$) are virially unbound.  Cores in the upper half of the virial plane ($\Omega_{\textsc{g}}/\Omega_{\textsc{p}}>1$) are gravitationally-dominated, while cores in the lower half of the virial plane ($\Omega_{\textsc{g}}/\Omega_{\textsc{p}}<1$) are external-pressure-dominated.  We model the loci of starless cores in this plane as a function of core characteristic radius.

\begin{figure}
\centering
\includegraphics[width=0.47\textwidth]{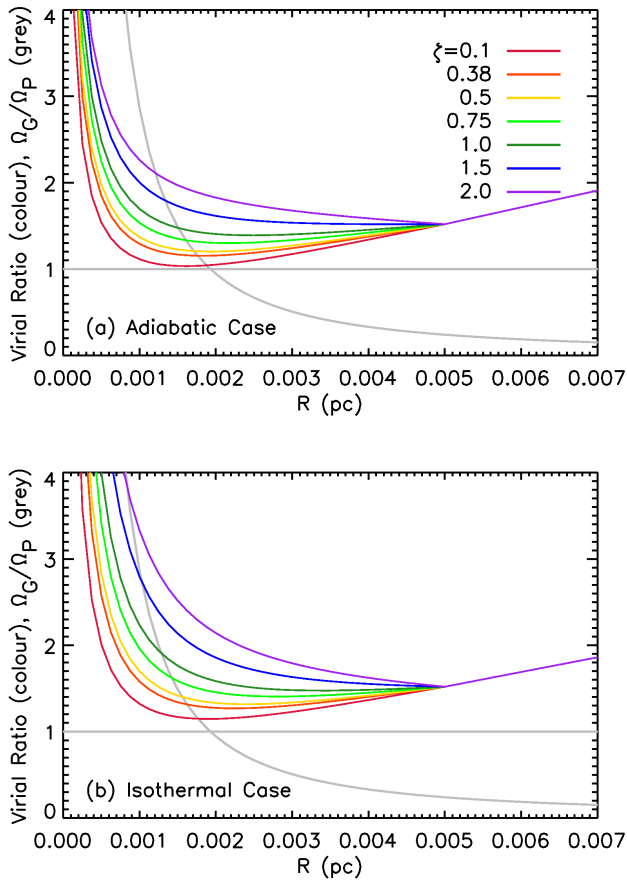}
\caption{Variation in behaviour of the virial ratio with index of relationship between core size and non-thermal linewidth for a Plummer-like core with $M=0.25\,$M\sun, $P_{\textsc{ext}}/k_{\textsc{b}}=1.5\times10^{7}$\,K\,cm$^{-2}$, $T_{0}=7\,$K, $\sigma_{\textsc{nt},0}=220\,$ms$^{-1}$ and $R_{0}=0.005\,$pc.  Top panel: adiabatic case.  Bottom panel: isothermal case.  Legend shows the index $\zeta$ where $\sigma_{\textsc{nt}}\propto r_{e}^{\zeta}$.  Note the similarity between the behavious resulting from the \citet{solomon1987} index of 0.5 (yellow) and the \citet{larson1981} index of 0.38 (orange).  The line of unity and the $\Omega_{\textsc{g}}/\Omega_{\textsc{p}}$ ratio are plotted in grey, for reference.}
\label{fig:vary_index}
\end{figure}

Figure~\ref{fig:evolution_example_1} shows the loci in the virial plane predicted by our model for a family of Plummer-like starless cores in the mass range 0.05--2.0\,M\sun\ with external pressure $P_{\textsc{ext}}/k_{\textsc{b}}=1.5\times10^{7}$\,K\,cm$^{-2}$, an initial temperature of 7\,K, an initial non-thermal linewidth of 250\,m\,s$^{-1}$, and an initial characteristic radius of 0.005\,pc.  Figure~\ref{fig:evolution_example_1}a shows the contracting tracks, with the adiabatic track shown as a solid line and the isothermal track shown as a dashed line.  The expanding tracks are shown on Figure~\ref{fig:evolution_example_1}b.  The two sets of tracks are separated for clarity.  We propose that each core will have an evolutionary track in this plane, along the locus defined by Equations~\ref{eq:density}--\ref{eq:sigmant_ev}.  As discussed above, we expect virially bound and pressure-confined ($-(\Omega_{\textsc{g}}+\Omega_{\textsc{p}})/2\Omega_{\textsc{k}}>1$ and $\Omega_{\textsc{g}}/\Omega_{\textsc{p}}<1$) cores to contract toward virial equilibrium.  Thus, for each of the cores shown in Figure~\ref{fig:evolution_example_1}, only some part of the locus of Equations~\ref{eq:density}--\ref{eq:sigmant_ev} is accessible, and represents an evolutionary track.  The 2.0, 1.0, 0.5 and 0.25\,M\sun\ cores, we expect to follow the contracting track indefinitely.  The 0.1\,M\sun\ core, we expect to follow the contracting track indefinitely in the isothermal case, and to contract to virial equilibrium in the adiabatic case.  We expect the 0.05\,M\sun\ core to follow the contracting track to virial equilibrium.  All of the cores in this family are initially virially bound; we expect them to contract, either to equilibrium or indefinitely.

Figure~\ref{fig:evolution_example_2} shows the loci in the virial plane we predict for a less-bound family of starless cores, in the mass range 0.01--0.5\,M\sun\ with external pressure $P_{\textsc{ext}}/k_{\textsc{b}}=1.5\times10^{7}$\,K\,cm$^{-2}$, an initial temperature of 7\,K, an initial non-thermal linewidth of 280\,m\,s$^{-1}$, and an initial characteristic radius of 0.002\,pc.  These cores show a more varied range of behaviours: the 0.5\,M\sun\ core is gravitationally and virially bound and will collapse indefinitely under gravity.  The 0.25, 0.1, 0.05 and 0.025\,M\sun\ cores are virially unbound; we expect these cores to follow the expanding track to virial equilibrium.  The 0.01\,M\sun\ core is virially bound and pressure-dominated; we expect this core to follow the contracting track to virial equilibrium.

\section{Non-thermal motions}
\label{sec:nonthermal}

\begin{figure}
\centering
\includegraphics[width=0.47\textwidth]{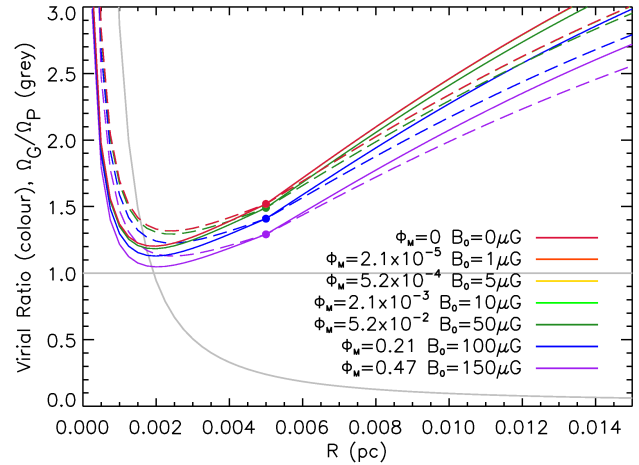}
\caption{Variation in behaviour of the virial ratio with initial magnetic field strength $B_{0}$, for a core with $M=0.25\,$M\sun, $P_{\textsc{ext}}/k_{\textsc{b}}=1.5\times10^{7}$\,K\,cm$^{-2}$, $T_{0}=7\,$K, $\sigma_{\textsc{nt},0}=250\,$ms$^{-1}$ and $R_{0}=0.005\,$pc.  Adiabatic curves are shown as solid lines; isothermal curves are shown as dashed lines.  The 0\,$\upmu$G, 1\,$\upmu$G, 5\,$\upmu$G and 10\,$\upmu$G curves overlap.  The line of unity and the $\Omega_{\textsc{g}}/\Omega_{\textsc{p}}$ ratio are plotted in grey, for reference.}
\label{fig:vary_bfield_basu2000}
\end{figure}

We choose to parameterise the non-thermal motions of our cores as Larson-like (i.e. $\sigma_{\textsc{nt}}\propto r_{e}^{\zeta}$ -- see Equation~\ref{eq:sigmant_ev}), in order to include the dissipation of turbulence expected to occur in starless cores (e.g., \citealt{klessen2005}; \citealt{offner2008}) in our model.

It is important to note that the scale-free, Kolmogorov-type, behaviour which is parameterised by a Larson-like size-linewidth relation may not apply on the smallest size scales which we consider here.  A number of different values have been determined for the turbulent energy dissipation scale in molecular clouds, the length scale below which turbulent motions dissipate rapidly.  \citet{ostriker2001} found a non-constant spectral index in velocity dispersion spectra created from magnetohydrodynamic simulations of turbulent giant molecular clouds.  They interpreted the change in spectral index at smaller size scales as being indicative of a turbulent energy dissipation scale.  \citet{ossenkopf2002} found a turbulent energy dissipation scale of 0.05\,pc in the Polaris Flare, larger than the typical core size considered in this work.  However, \citet{offner2008} found Larson-like dissipation of turbulence continuing to scales $\sim0.01$\,pc in their simulations of molecular clouds.  \citet{li2008} found that HCN observations of M17 showed a constant spectral index above size scales $\sim 0.16$\,pc (their limiting spatial resolution), and inferred a turbulent energy dissipation scale of 0.0018\,pc, similar to the smaller size scales considered in this work.  \citet{pattle2015} found core non-thermal velocity dispersions in the Ophiuchus molecular cloud to be typically mildly supersonic.  While this does not allow determination of the turbulent energy dissipation scale in this region, it suggests that the turbulent energy dissipation scale is smaller than the measured characteristic radii of cores in this region ($\sim 0.0015-0.0065$\,pc; see Table~\ref{tab:ics}, below), and that Larson-like behaviour may be valid for the cores considered in this work at least on larger scales.

Larson-like behaviour is not known to hold at the smallest scales, and at these smallest scales our values of the the internal non-thermal linewidths of the cores may be overestimated.  However, in the absence of a means of determining the turbulent energy dissipation scale for the cores we consider, we parameterise the dissipation of turbulence as Larson-like on all scales smaller than the initial size of the core.

We choose an index $\zeta=0.5$ \citep{solomon1987}.  However, other indices have been proposed -- for example, for whole molecular clouds an index of $\zeta=0.38$ is expected \citep{larson1981}, while \citet{caselli1995} find an index $\zeta=0.21$ in high-mass star-forming regions.  We investigated how our evolutionary model varies with $\zeta$.  We found that as the non-thermal contribution to the core's kinetic energy becomes small at small radii, and as the dependence of the non-thermal linewidth on size is relatively weak ($\sigma_{\textsc{nt}}\propto r_{e}^{0.5}$), over the range of radii being considered, small changes in the index of Equation~\ref{eq:sigmant_ev} do not substantially alter the expected behaviour of our cores.

While $\sigma_{\textsc{nt}}\propto r_{e}^{\zeta}$ and $0<\zeta<1$, the behaviour of our cores does not alter significantly with varying $\zeta$.  An $\zeta=0$ would indicate that there is no dissipation of turbulence as the core contracts, while $\zeta<0$ would require turbulence to be enhanced, rather than dissipated, as the core decreases in size.  A value of $\zeta<0.5$ implies a sub-linear increase in non-thermal kinetic energy as a function of $r_{e}$ (as $\Omega_{\textsc{k,nt}}\propto\sigma_{\textsc{nt}}^{2}$).  If $\zeta>0.5$, then the substantial increase in non-thermal kinetic energy with increasing $r_{e}$ that this causes begins to destroy the minimum in virial ratio seen in the intermediate-$R$ region of Figure~\ref{fig:hypothetical}.  As discussed above, we assume that $\sigma_{\textsc{nt}}(R>R_{0})=\sigma_{\textsc{nt}}(R_{0})$, and so the core's non-thermal kinetic energy does not increase as a core expands from its initial size.

The variation in the virial ratio with $\zeta$ is shown in Figure~\ref{fig:vary_index}, for a core with mass $M=0.25$\,M\sun, external pressure $P_{\textsc{ext}}/k_{\textsc{b}}=1.5\times10^{7}$\,K\,cm$^{-3}$, external density $\rho_{e}=10^{5}$ H$_{2}$\,molecules\,cm$^{-3}$, and, at an initial characteristic radius of 0.005\,pc, a temperature of 7\,K and a non-thermal internal velocity dispersion of 220\,ms$^{-1}$.  Figure~\ref{fig:vary_index} shows that as $\zeta$ increases, the virial ratio of the core increases and the virial minimum at small radii becomes less distinct; i.e. the core is dissipating turbulence more effectively.  Figure~\ref{fig:vary_index} also shows that there is only a small difference in the behaviour of the virial ratio between the \citet{solomon1987} index of 0.5 and the \citet{larson1981} index of 0.38.

We choose the \citet{solomon1987} value for the index of the relation between characteristic radius and non-thermal linewidth (i.e. $\zeta=0.5$) as being justifiable and physically plausible, while noting that varying this value within a physically reasonable range would not substantially alter our results.

\section{Parameterisation of magnetic field term}
\label{sec:magnetism}

We have thus far neglected the magnetic field term in the virial equation in this analysis.  However, we note that the effect of the magnetic field on the virial balance of a core can be included in our model.

Assuming that the \citet{basu2000} relation,
\begin{equation}
B\propto n^{\nicefrac{1}{2}}\sigma_{\textsc{nt}},
\label{eq:basu}
\end{equation}
(where $B$ is magnetic field strength and $n$ is number density) holds for our starless cores, then, as shown by \citet{pattle2015}, there is a constant ratio between magnetic energy and non-thermal kinetic energy:
\begin{equation}
\frac{\Omega_{\textsc{m}}}{\Omega_{\textsc{k,nt}}}=\frac{1}{3\mu_{0}}\frac{B_{0}^{2}}{\rho_{0}\sigma_{0,\textsc{nt}}^{2}}=\Psi_{\textsc{m}}.
\label{eq:b_nt}
\end{equation}
$\Psi_{\textsc{m}}=B_{0}^{2}/3\mu_{0}\rho_{0}\sigma_{0,\textsc{nt}}^{2}$ is the ratio of magnetic energy to non-thermal kinetic energy, from measurement of a magnetic field strength $B_{0}$ and a non-thermal linewidth $\sigma_{0,\textsc{nt}}$ in material with a density $\rho_{0}$ ($\mu_{0}$ is the permeability of free space).

It is important to note that the relation given in equation~\ref{eq:basu} was determined for flattened, disc-like structures, rather than the spherical geometries which we consider in this work.  We use equation~\ref{eq:b_nt} as a convenient means by which to parameterise the effect of internal magnetic field on our cores, while noting that the applicability of equation~\ref{eq:basu} to this problem is not certain.

Including the magnetic term, the virial ratio becomes
\begin{equation}
{\rm Virial\,\,Ratio} = -\frac{\Omega_{\textsc{g}}+\Omega_{\textsc{p}}}{2\Omega_{\textsc{k}}+\Omega_{\textsc{m}}},
\end{equation}
and, if the \citet{basu2000} relation holds and $\Omega_{\textsc{m}}=\Psi_{\textsc{m}}\Omega_{\textsc{k,nt}}$, then
\begin{equation}
-\frac{\Omega_{\textsc{g}}+\Omega_{\textsc{p}}}{2\Omega_{\textsc{k}}+\Omega_{\textsc{m}}}=-\frac{\Omega_{\textsc{g}}+\Omega_{\textsc{p}}}{2\Omega_{\textsc{k,t}}+(2+\Psi_{\textsc{m}})\Omega_{\textsc{k,nt}}},
\end{equation}
and we can continue to model the evolution of our cores as a function of characteristic radius $R$ only, although another initial condition, initial magnetic field strength $B_{0}$, is now required.

The variation in the virial ratio with $\Psi_{\textsc{m}}$ is shown in Figure~\ref{fig:vary_bfield_basu2000}, again for a core with mass $M=0.25$\,M\sun, external pressure $P_{\textsc{ext}}/k_{\textsc{b}}=1.5\times10^{7}$\,K\,cm$^{-3}$, external density $\rho_{e}=10^{5}$ H$_{2}$\,molecules\,cm$^{-3}$, and a temperature of 7\,K and a non-thermal internal velocity dispersion of 250\,ms$^{-1}$ at a characteristic radius of 0.005\,pc.  We estimate $\Psi_{\textsc{m}}$ for a representative initial core density of $\rho_{0}=3M(r_{e,0})/4\uppi r_{e,0}^{3}$ and a range of magnetic field strengths $B_{0}$.

Figure~\ref{fig:vary_bfield_basu2000} shows that for the chosen set of initial conditions, the cases of $B_{0}=0\,\upmu$G $(\Psi_{\textsc{m}}=0)$, $B_{0}=1\,\upmu$G $(\Psi_{\textsc{m}}=2.1\times10^{-5})$, $B_{0}=5\,\upmu$G $(\Psi_{\textsc{m}}=5.2\times10^{-4})$ and $B_{0}=10\,\upmu$G $(\Psi_{\textsc{m}}=0.0021)$ are not distinguishable; the contribution of the magnetic field to the energy balance of the core is negligible.  In the case $B_{0}=50\,\upmu$G $(\Psi_{\textsc{m}}=0.052)$, the effect of the magnetic energy term is visible on Figure~\ref{fig:vary_bfield_basu2000}, but not sufficient to cause more than a minimal variation in the core's evolutionary track.  In the case of this core, it is not until field strengths such as $B_{0}=100\,\upmu$G $(\Psi_{\textsc{m}}=0.21)$ are reached that the energy balance begins to change significantly.

\section{Application to cores in L1688}
\label{sec:ophiuchus}

\begin{table*}
\centering
\caption{Initial conditions for cores in Ophiuchus.  Measured values of mass, characteristic radius, non-thermal velocity dispersion and external pressure are taken from from \citet{pattle2015}.  The bounding radius is calculated from measured properties using equations~\ref{eq:re_p} and \ref{eq:mass_trunc_p}.}
\label{tab:ics}
\begin{tabular}{c c c c c c c}
\toprule
Source & $M$ & $R_{0}$ & $\sigma_{\textsc{nt},0}$ & $P_{\textsc{ext}}/k_{\textsc{b}}$ & $r_{e,0}$ & \multirow{2}{*}{$\dfrac{r_{e,0}}{R_{0}}$}\\
ID & (M$_{\odot}$) & (pc) & (m\,s$^{-1}$) & ($\times 10^{7}$\,K\,cm$^{-3}$) & (pc) & \\
\midrule
SM1 & 1.30 & 0.0033 & 270 & 0.79 & 0.0166 & 5.06 \\
SM1N & 1.00 & 0.0029 & 266 & 1.11 & 0.0151 & 5.15 \\
SM2 & 0.76 & 0.0050 & 197 & 0.75 & 0.0164 & 3.29 \\
A-MM5 & 0.26 & 0.0061 & 216 & 0.79 & 0.0134 & 2.20 \\
A-MM6 & 0.75 & 0.0063 & 245 & 0.86 & 0.0175 & 2.78 \\
A-MM7 & 0.26 & 0.0053 & 259 & 0.75 & 0.0129 & 2.43 \\
A-MM8 & 0.28 & 0.0048 & 166 & 0.47 & 0.0127 & 2.64 \\
A-MM4 & 0.11 & 0.0048 & 179 & 0.53 & 0.0102 & 2.12 \\
A-MM4a & 0.04 & 0.0015 & 173 & 0.60 & 0.0058 & 3.96 \\
B1-MM3 & 0.27 & 0.0051 & 174 & 1.06 & 0.0129 & 2.52 \\
B1-MM4a & 0.29 & 0.0050 & 232 & 1.04 & 0.0131 & 2.61 \\
B1-MM4b & 0.06 & 0.0021 & 200 & 1.05 & 0.0070 & 3.34 \\
B2-MM6 & 0.56 & 0.0056 & 358 & 1.53 & 0.0158 & 2.80 \\
B2-MM9 & 0.61 & 0.0065 & 286 & 1.54 & 0.0167 & 2.58 \\
B2-MM13 & 0.62 & 0.0050 & 307 & 1.09 & 0.0156 & 3.15 \\
B2-MM14 & 0.79 & 0.0064 & 251 & 1.68 & 0.0178 & 2.77 \\
B2-MM15 & 0.35 & 0.0043 & 225 & 2.38 & 0.0130 & 3.04 \\
B2-MM16 & 0.25 & 0.0015 & 276 & 0.84 & 0.0089 & 6.11 \\
C-MM3 & 0.24 & 0.0054 & 158 & 1.98 & 0.0128 & 2.35 \\
C-MM6a & 0.08 & 0.0035 & 164 & 2.25 & 0.0086 & 2.43 \\
C-MM6b & 0.09 & 0.0053 & 165 & 2.20 & 0.0101 & 1.89 \\
E-MM2d & 0.15 & 0.0045 & 124 & 1.16 & 0.0108 & 2.40 \\
F-MM1 & 0.05 & 0.0015 & 153 & 1.72 & 0.0062 & 4.22 \\
\bottomrule
\end{tabular}
\end{table*}

We apply this model to the 23 starless cores in the L1688 region of Ophiuchus on which \citet{pattle2015} performed a full virial analysis.  \citet{pattle2015} determined core masses from SCUBA-2 and Herschel flux density measurements taken as part of the JCMT Gould Belt Survey \citep{wardthompson2007} and Herschel Gould Belt Survey \citep{andre2010} respectively, internal linewidths from IRAM N$_{2}$H$^{+}$ $1\to 0$ data originally presented by \citet{andre2007}, and external linewidths from HARP C$^{18}$O $3\to 2$ measurements originally presented by \citet{white2015}.  In this section, we predict evolutionary outcomes for these cores, taking the core properties given by \citet{pattle2015} as our set of initial conditions.  These initial conditions are listed in Table~\ref{tab:ics}.

We neglect the magnetic energy term in the virial equation in the following analysis, due to the uncertainty of the applicability of the magnetic energy analysis presented above to the problem.  In Ophiuchus, \citet{pattle2015} determined a value of $\Psi_{\textsc{m}}=0.11$, based on measurements by \citet{troland1996}.  This is in the range which will produce a small change in the predicted evolutionary track of the core, but will change the predicted evolutionary outcome of the core only in the most marginal cases.  This suggests that our neglect of the the magnetic energy term in Ophiuchus is justifiable.

\begin{figure*}
\centering
\includegraphics[width=0.9\textwidth]{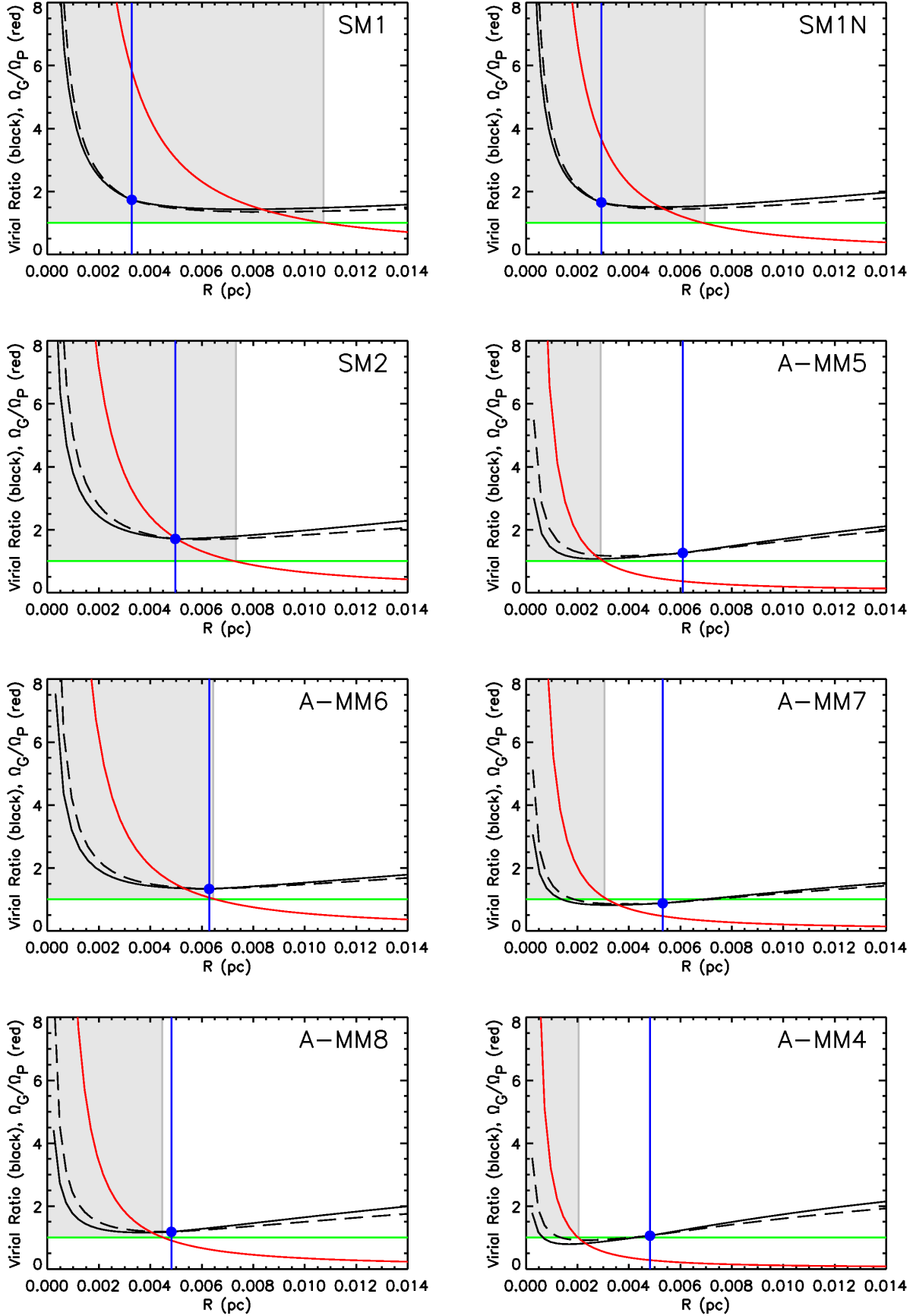}
\caption{Stability as a function of characteristic radius for our cores.  Solid black line shows adiabatic virial ratio; dashed black line shows isothermal virial ratio.  Red line shows ratio of gravitational potential energy to external pressure energy. -- \emph{cont'd overleaf}}
\label{fig:oph_stability}
\end{figure*}
\addtocounter{figure}{-1}
\begin{figure*}
\centering
\includegraphics[width=0.9\textwidth]{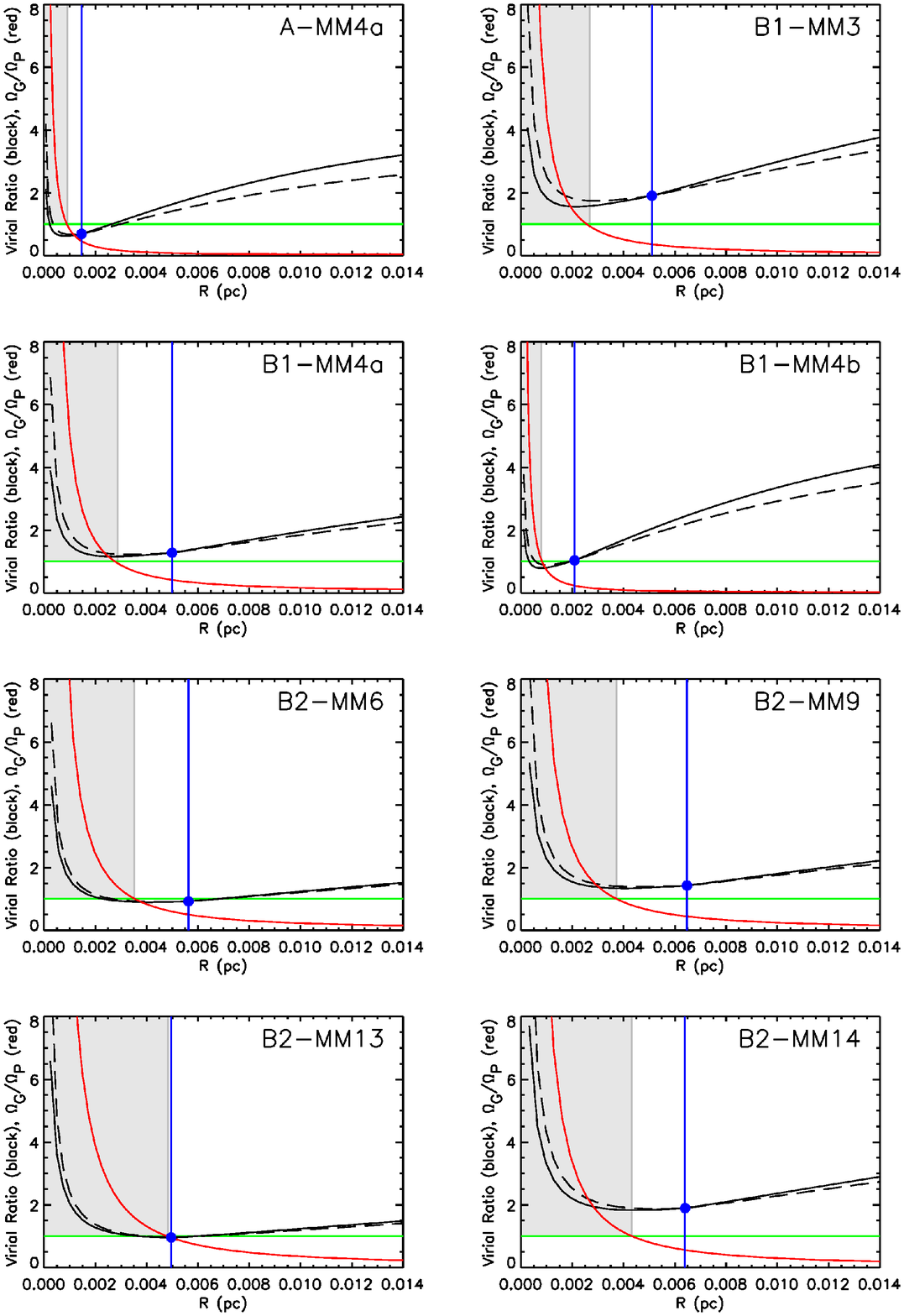}
\caption{-- \emph{cont'd.}  Blue dot shows measured virial ratio (blue line is present to guide the eye to the measured confinement ratio).  Green line shows line of virial stability.  Grey shaded regions indicate parameter space occupied by prestellar cores.}
\end{figure*}
\addtocounter{figure}{-1}
\begin{figure*}
\centering
\includegraphics[width=0.9\textwidth]{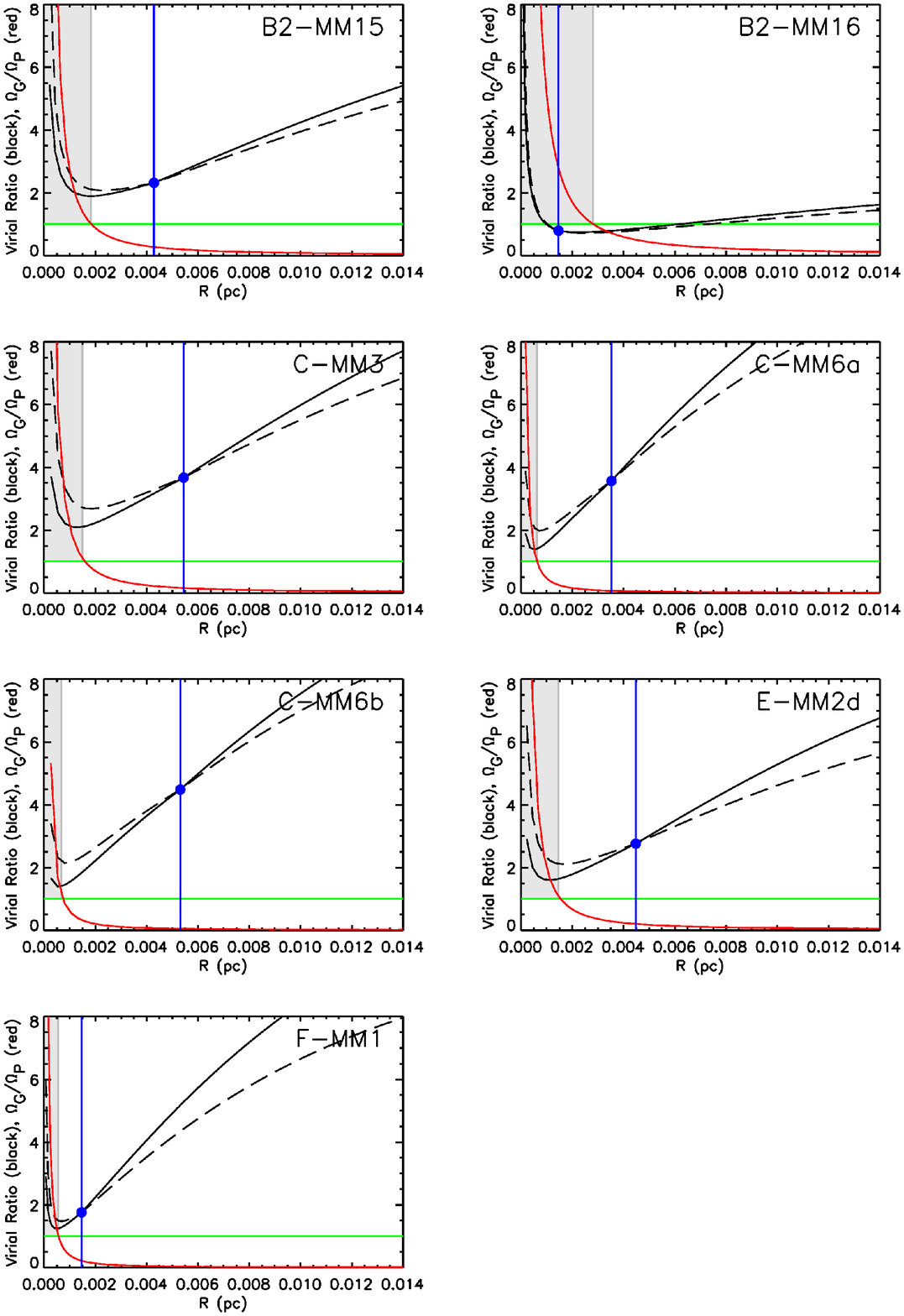}
\caption{-- \emph{cont'd.}}
\end{figure*}

The model we use in this paper differs somewhat from that used by \citet{pattle2015} to assess the virial stability of their cores.  \citet{pattle2015} assumed that their cores obeyed Gaussian density distributions and took the gravitational potential energies of their cores to be those of infinite Gaussian density distributions of the same total mass.  In this work, we assume that the masses $M$ measured by \citet{pattle2015} and listed in Table~\ref{tab:ics} represent the masses enclosed in the cores' bounding radii, $r_{e}$ (where $r_{e}$ is calculated from mass, characteristic radius and bounding density; see equations~\ref{eq:re_p} and \ref{eq:mass_trunc_p}) , and that the cores obey a Plummer-like density distribution with characteristic radius $R$ equal to the Gaussian width measured by \citet{pattle2015}.  This results in our initial values of the virial and confinement ratios being slightly different from those given by \citet{pattle2015}.  However, the difference in the model typically produces a difference in virial and confinement ratios smaller than the uncertainties quoted by \citet{pattle2015}.  We reproduce the virial plane determined by \citet{pattle2015} in Appendix C, for purposes of comparison.

\begin{table*}
\centering
\caption{The evolutionary outcomes predicted for each of our cores in Ophiuchus by the analytical model presented here.}
\begin{tabular}{c c c c cc}
\bottomrule
 & Initial & Confining & Direction of & \multicolumn{2}{c}{Predicted Outcome} \\ \cmidrule{5-6}
Core & State & Force & Evolution & Isothermal & Adiabatic \\
\midrule
SM1 & Bound & Gravity & Contraction & Prestellar & Prestellar \\
SM1N & Bound & Gravity & Contraction & Prestellar & Prestellar \\
SM2 & Bound & Gravity & Contraction & Prestellar & Prestellar \\
A-MM5 & Bound & Pressure & Contraction & Prestellar & Prestellar \\
A-MM6 & Bound & Gravity & Contraction & Prestellar & Prestellar \\
A-MM7 & Unbound & Pressure & Expansion & Virialised & Virialised \\
A-MM8 & Bound & Pressure & Contraction & Prestellar & Prestellar \\
A-MM4 & Bound & Pressure & Contraction & Virialised & Virialised \\
A-MM4a & Unbound & Pressure & Expansion & Virialised & Virialised \\
B1-MM3 & Bound & Pressure & Contraction & Prestellar & Prestellar \\
B1-MM4a & Bound & Pressure & Contraction & Prestellar & Prestellar \\
B1-MM4b & Bound & Pressure & Contraction & Virialised & Virialised \\
B2-MM6 & Unbound & Pressure & Expansion & Virialised & Virialised \\
B2-MM9 & Bound & Pressure & Contraction & Prestellar & Prestellar \\
B2-MM13 & Unbound & Pressure & Expansion & Virialised & Virialised \\
B2-MM14 & Bound & Pressure & Contraction & Prestellar & Prestellar \\
B2-MM15 & Bound & Pressure & Contraction & Prestellar & Prestellar \\
B2-MM16 & Unbound & Gravity & Expansion & Virialised &  Virialised \\
C-MM3 & Bound & Pressure & Contraction & Prestellar & Prestellar \\
C-MM6a & Bound & Pressure & Contraction & Prestellar & Prestellar \\
C-MM6b & Bound & Pressure & Contraction & Prestellar & Prestellar \\
E-MM2d & Bound & Pressure & Contraction & Prestellar & Prestellar \\
F-MM1 & Bound & Pressure & Contraction & Prestellar & Prestellar \\
\bottomrule
\end{tabular}
\label{tab:evolutionary_outcomes} 
\end{table*}

Figure~\ref{fig:oph_stability} shows the virial stability as a function of characteristic radius of each of the starless cores in L1688, plotted in the manner described for Figure~\ref{fig:hypothetical}.  Each panel of Figure~\ref{fig:oph_stability} shows, for an individual starless core in our sample, the virial ratio, $-(\Omega_{\textsc{g}}+\Omega_{\textsc{p}})/2\Omega_{\textsc{k}}$, in black and the confinement ratio, $\Omega_{\textsc{g}}/\Omega_{\textsc{p}}$, in red, both plotted as a function of core characteristic radius $R$.

We again assume an external density $\rho_{e}=10^{5}$ H$_{2}$\, molecules\,cm$^{-3}$ and a mean molecular mass $2.86\times m_{\textsc{h}}$.  \citet{pattle2015} assumed this bounding density to be representative of the density above which C$^{18}$O ceases to be an effective tracer of core material, becoming significantly depleted onto dust grains, and also to be representative of the density at which \nh\ becomes detectable \citep{difrancesco2007}.  \citet{pattle2015} determined masses for the cores which they detected in Ophiuchus using dust continuum emission, \nh\ emission and \co\ emission, and found that masses determined from \nh\ correlated well with masses determined from dust emission, while the correlation between \co-detemined masses and dust-emission-determined masses was much weaker, which they interpreted as indicating that \nh\ was tracing the dense gas in the cores, while \co\ was tracing the somewhat less dense gas surrounding the core.  As the continuum emission and \nh\ emission were interpreted to be tracing the same material, the minimum density at which \nh\ is detectable was taken to be representative of the density bounding the cores, whose sizes were measured from continuum emission.  As the main route to destruction of the \nh\ molecule in dense environments is the reaction 
\begin{equation}
{\rm N}_{2}{\rm H}^{+}+{\rm CO}\to {\rm N}_{2}+{\rm HCO}^{+},
\end{equation}
(e.g. \citealt{snyder1977}, and references therein), \nh\ can only reach high abundances at high densities, when CO (and its isotopologues) is not present in the gas phase.  This makes \nh\ and \co\ unlikely to be tracing the same material.  An example of this is seen in the CO `snow line' in the protostellar disc surrounding the star HD163296 \citep{qi2015}.  \citet{qi2015} showed that CO traces warmer disc material near to the protostar, and that \nh\ traces a ring of cooler material at larger radii.  \citet{qi2015} further showed that that the radius of the CO snow line corresponds well with the radius at which \nh\ emission becomes detectable, indicating that the two molecules are tracing two separate but contiguous regions of the disc.  \citet{pattle2015} considered the inverse situation in Ophiuchus, in which \nh\ traces the cool and dense starless core material, while \co\ traces the warmer and more rarefied material immediately surrounding the cores.

According to our model, there are four gravitationally-dominated and virially-bound cores in our sample: SM1, SM1N, SM2 and A-MM6.  We assume that these cores are prestellar and collapsing under gravity, and will evolve away from virial equilibrium.

Of the pressure-confined and virially-bound starless cores in our sample, we expect A-MM5, A-MM8, B1-MM3, B1-MM4a, B2-MM9, B2-MM14, B2-MM15, C-MM3, C-MM6a, C-MM6b, E-MM2d and F-MM1 to evolve into gravitationally-bound prestellar cores.  A-MM4 and B1-MM4b, we expect to contract to virial equilibrium.

Cores A-MM7, A-MM4a, B2-MM6 and B2-MM13 are virially unbound and pressure-dominated, and will expand to reach virial equilibrium.

The one gravitationally-dominated and virially-unbound starless core in our sample is B2-MM16.  We predict that this core will expand to virial equilibrium, despite this initially increasing the core's virial instability.  We note that the uncertainty on the virial ratio of this core given by \citet{pattle2015} is large enough for B2-MM16 to be consistent with in fact being a gravitationally-bound prestellar core.  We discuss this core further below.

The predicted evolutionary outcomes of our cores are listed in Table~\ref{tab:evolutionary_outcomes}.  We emphasise that all of these evolutionary outcomes assume that there is no further accretion of mass by the core.  We address this in a future paper.

\begin{figure}
\centering
\includegraphics[width=0.47\textwidth]{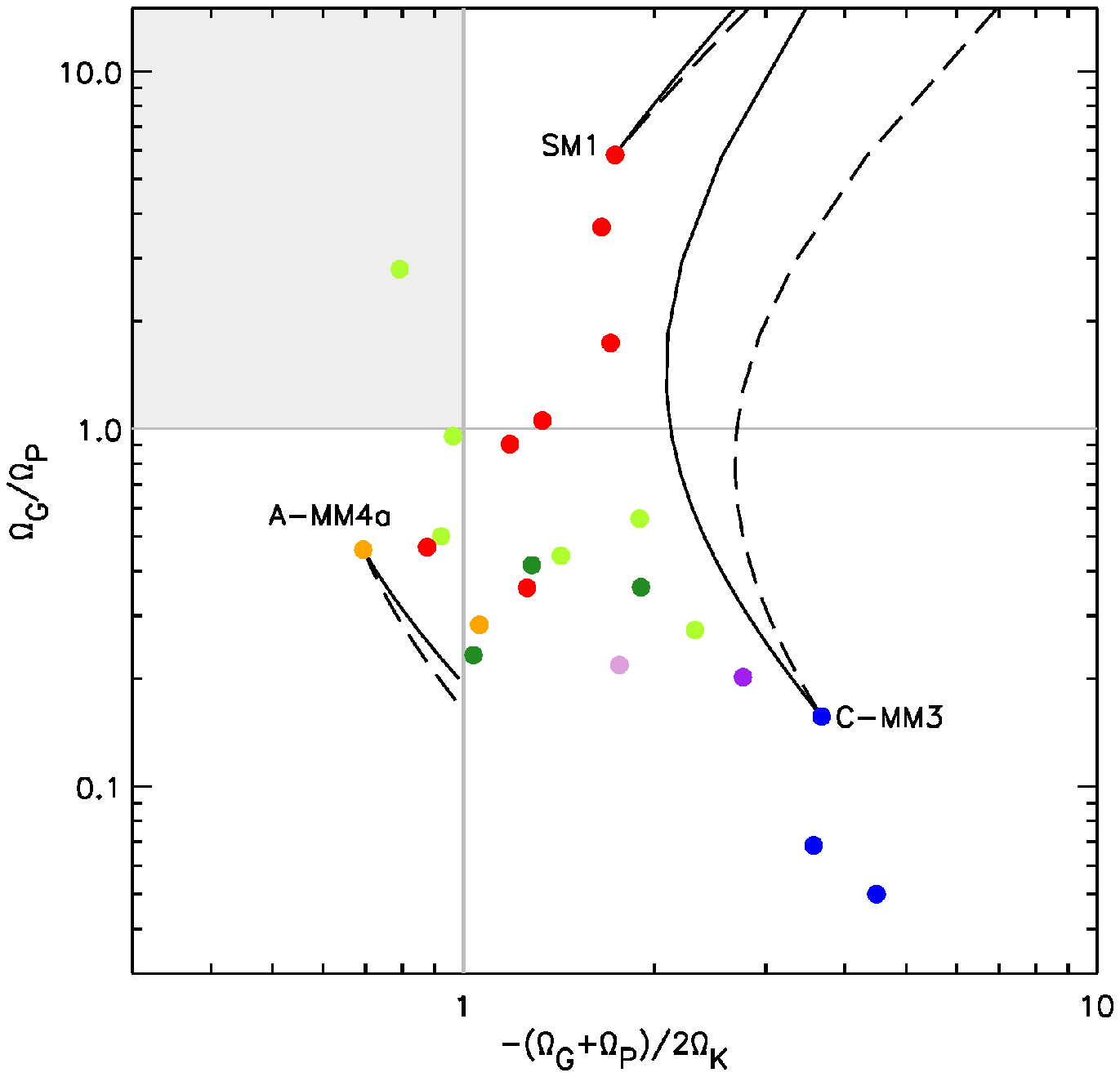}
\caption{Evolutionary tracks in the virial plane, for a sample of the \citet{pattle2015} starless cores.  All cores shown in Figure~\ref{fig:oph_stability} are shown here for reference.  See text for details.  Colour coding indicates region, as defined by \citet{motte1998}: red, Oph A; orange, Oph A$^{\prime}$ (low-column-density region surrounding Oph A); dark green, Oph B1; light green, Oph B2; blue, Oph C; dark purple, Oph E; light purple, Oph F.  Solid black line indicates adiabatic track; dashed back line indicates isothermal track.  The shaded region indicates the hypothesised `starless core desert'.}
\label{fig:evolution}
\end{figure}

Figure~\ref{fig:evolution} shows our proposed evolutionary tracks for a subset of our cores: SM1 (prestellar), C-MM3 (pressure-confined, contracting to prestellar), and A-MM4a (unbound and pressure-dominated, expanding to virialised).  These are chosen to illustrate the behaviours described above.

\section{Discussion}

\subsection{A `starless core desert'?}

It is notable that there is only one core in the L1688 region which is both gravitationally dominated and virially unbound, and that this one core, B2-MM16, has an uncertainty on its virial ratio such that the core is consistent with in fact being prestellar.  We hypothesise that this parameter space ($-(\Omega_{\textsc{g}}+\Omega_{\textsc{p}})/2\Omega_{\textsc{k}}<1$; $\Omega_{\textsc{g}}>\Omega_{\textsc{p}}$) may be largely inaccessible to starless cores - a `starless core desert' of sorts.  It may be difficult to assemble a starless core with sufficient mass to be gravitationally-dominated, while simultaneously maintaining an internal linewidth sufficiently large that the core remains virially unbound.

The absence of virially-unbound and gravitationally-dominated starless cores further hints at a formation mechanism for prestellar cores in which starless cores initially form as pressure-confined objects, and those which are sufficently virially bound then evolve to become gravitationally-bound prestellar cores, perhaps in the manner described in this model.

In order to test this hypothesis, further measurements of the virial balance of starless cores are needed.  If the gravitationally-dominated and virially-unbound quadrant of the virial plane is in fact significantly underpopulated compared to the other quadrants -- i.e. if the hypothesised `starless core desert' is not a result of the small counting statistics in the L1688 core sample -- this would lend support to the suggestion that starless cores may initially form as pressure-confined objects.

\subsection{Comparison with numerical modelling}

This model can be usefully compared to numerical simulations of cores collapsing under external pressure.  Such simulations typically assume that cores obey a Bonnor-Ebert density distribution, and typically involve the perturbation of a system which is initially in equilibrium  (e.g. \citealt{foster1993}; \citealt{hennebelle2003}).  This is a somewhat different approach to our model, which considers cores as obeying a Plummer-like density distribution, and models the evolution of cores which are initially in a non-equilibrium (i.e. virially unstable) state.

\citet{hennebelle2003} modelled the evolution of an initially stable Bonnor-Ebert sphere undergoing a steady increase in external pressure, in order to study protostellar collapse induced by external compression.  They found that while the compression of their core is slow (i.e. when the external pressure on their core increases on a timescale much greater than the sound-crossing time of the core), the core evolves quasistatically.  During the prestellar stage of the core's evolution the outer boundary of the core is pushed inward -- qualitatively similar to the contraction of pressure-dominated and virially bound cores in our model -- and a modest, approximately uniform, inward velocity field is set up.  However, \citet{hennebelle2003} found that when cores are strongly compressed (i.e. the external pressure increases on a timescale shorter than the sound-crossing time), a compression wave is driven into the core, leaving behind it an inward velocity field which can become supersonic if the core compression is strong enough.  This is dissimilar to our model, which assumes quasistatic core evolution throughout.

Our model is thus qualitatively similar to numerical simulations of the collapse under slow compression of pressure-confined cores \citep{hennebelle2003}.  Whether the environments in the molecular clouds studied in this work allow quasistatic core evolution is not clear.  However, \citet{hennebelle2003} noted that their simulations in which core compression is slow -- the quasistatic case -- produce results which match observational constraints on starless cores, suggesting that core evolution may be quasistatic in at least some cases.

\subsection{The definition of `prestellar'}

This analysis shows that a virially-bound and pressure-confined starless core will not necessarily evolve to become gravitationally bound, and thus cannot be considered to be a prestellar core.  Those of our cores which have no route to becoming gravitationally bound may be evolving toward or oscillating slightly about virial equilibrium.  A core can only be definitively considered prestellar (i.e. about to form a protostar) if it is gravitationally unstable.  Pressure confinement alone is not necessarily sufficient.

\subsection{Observational uncertainties}

It must be emphasised that the core properties measured by \citet{pattle2015} and listed as initial conditions in Table~\ref{tab:ics} are substantially uncertain.  The majority of the cores have virial ratios consistent at the 3-$\sigma$ level with their being virialised, and the evolutionary tracks described above are accurate only if measurements of the core properties are precisely accurate.  These evolutionary scenarios should be viewed as representative of a core with the described properties, rather than a prediction specific to the core being observed.

\section{Summary}

In this paper we have presented an analytic model for the evolution of a starless core, based on seven observable quantities: mass, size, temperature, internal velocity dispersion, external velocity dispersion, external density and magnetic field strength.  This model assumes that starless cores obey a monotonically decreasing density distribution, truncated at a constant external density and confined by a constant external pressure.  We considered Plummer-like and Gaussian density distributions.  Core evolution was considered in the isothermal and adiabatic limits.  Non-thermal internal motions were parameterised as Larson-like, obeying the \citet{solomon1987} $\sigma_{\textsc{nt}}\propto r_{e}^{0.5}$ relation.  The magnetic energy of the core was parameterised as proportional to the non-thermal internal kinetic energy of the core, a result derived from the \citet{basu2000} relation between magnetic field strength and number density. 

 We modelled the virial ratio and the ratio of gravitational potential energy to external pressure energy of a core as a function of core characteristic radius.  We constructed evolutionary tracks in the `virial plane' diagram introduced by \citet{pattle2015}, under the assumption that a gravitationally unstable and virially bound core will collapse under gravity away from virial equilibrium, and that all other cores will expand or contract toward virial equilibrium as their energy balance dictates.  We found that not all pressure-confined and virially bound starless cores will evolve to become prestellar, with many contracting to equilibrium with their surroundings rather than becoming gravitationally unstable.  Therefore, we consider a core as prestellar -- i.e. in the process of collapsing to form a system of stars -- only if it is both virially unstable and its gravitational potential energy exceeds its energy from external pressure.

We considered the differences between adiabatic and isothermal contraction, and noted that cores contract more effectively under the assumption of isothermal contraction than under adiabatic contraction.  However, we note that our model is not physically justified at the extremely small characteristic radii at which the adiabatic and isothermal virial ratios differ significantly.  We discussed the variation in the virial ratio introduced by varying the index of the Larson relation, and found that varying the Larson index in a physically reasonable range does not substantially alter our results.  We discussed the magnetic field strengths necessary to alter the energy balance of a starless core, and found that the magnetic energy must be $\gtrsim 10$\% of the non-thermal kinetic energy for its contribution to the energy balance to be appreciable, and $\gtrsim 25$\% to significantly alter the energy balance of the core.

We applied our analysis to the 23 starless cores in the L1688 region of Ophiuchus for which \citet{pattle2015} determined an energy balance, assuming that the cores obey a Plummer-like density profile and are not magnetically-dominated.  We found that whether virially-bound and pressure-confined starless cores in L1688 will contract to virial equilibrium or gravitational instability depends sensitively on their measured properties.  We found that no more than one core in the L1688 region is both gravitationally-dominated and virially-unbound, and hypothesised a `starless core desert' in this quadrant of the virial plane.  We suggest that this parameter space may be inaccessible if starless cores initially form as pressure-confined objects. 

The model we present in this paper parameterises much of the physics of a starless core in terms of six (or seven, in the magnetic case) observable quantities.  It is envisaged as a means by which the likely evolutionary outcome of an observed starless core can be rapidly assessed without the need to perform detailed and computationally expensive simulations.

Throughout this analysis we have assumed that the core under consideration does not continue to accrete mass.  In a subsequent paper, we will consider the case in which the core's mass can vary.

\section{Acknowledgements}

The author wishes to thank Derek Ward-Thompson for valuable discussions and the anonymous referee for a detailed and constructive review which greatly improved the content of this paper, and STFC for studentship support under grant number ST/K501943/1 and postdoctoral support under grant number ST/K002023/1 while this research was carried out.

\bibliographystyle{mn2e_fix}

\section*{APPENDIX A: Gravitational Potential Energy of a Truncated Plummer-like Distribution}

\setcounter{equation}{0}
\makeatletter 
\renewcommand{\theequation}{A\@arabic\c@equation}
\makeatother

We give here a brief derivation of the gravitational potential energy of an $\eta=4$ Plummer-like distribution.

For a radially-symmetric potential, the gravitational potential energy $\Omega_{\textsc{g}}$ is given by
\begin{equation}
\Omega_{\textsc{g}}(r)=-4\uppi G\int_{0}^{r}{\rm d}r^{\prime}\,r^{\prime}\,\rho(r^{\prime})M(r^{\prime}), 
\end{equation}
where $\rho(r)$ and $M(r)$ are the density at and mass enclosed at radius $r$, respectively.  $M(r)$ is given by
\begin{equation}
M(r)=4\uppi\int_{0}^{r}{\rm d}r^{\prime}\,r^{\prime2}\rho(r^{\prime}). 
\end{equation}

We assume a radially-symmetric Plummer-like density distribution,
\begin{equation}
\rho(r)=\rho_{c}\left(\frac{R}{\sqrt{r^{2}+R^{2}}}\right)^{\eta}, 
\end{equation}
and choose an index $\eta=4$.
The total mass enclosed at radius $r$ is given by
\begin{equation}
M(r)=2\uppi\rho_{c}R^{3}\left[\arctan\left(\frac{r}{R}\right)-\frac{rR}{r^{2}+R^{2}}\right]. 
\end{equation}
and the total mass summed over all radii is given by
\begin{align}
M_{\rm inf} = \uppi^{2}\rho_{c}R^{3}. 
\end{align}

Using equations A1, A3 and A4, $\Omega_{\textsc{g}}(r)$ is given by
\begin{multline}
\Omega_{\textsc{g}}(r) = -8\uppi^{2}G\rho_{c}^{2}R^{7}\times \\
 \int^{r}_{0}{\rm d}r^{\prime}\frac{r^{\prime}}{(r^{\prime2}+R^{2})^{2}}\left[\arctan\left(\frac{r^{\prime}}{R}\right)-\frac{r^{\prime}R}{r^{\prime2}+R^{2}}\right]. 
\end{multline}
The first term in equation A6 is
\begin{multline}
\int^{r}_{0}{\rm d}r^{\prime}\frac{r^{\prime}}{(r^{\prime2}+R^{2})^{2}}\arctan\left(\frac{r^{\prime}}{R}\right) = \\
\frac{1}{4R^{2}}\left[\arctan\left(\frac{r}{R}\right)+\frac{rR}{r^{2}+R^{2}}\right]-\frac{\arctan\left(\frac{r}{R}\right)}{2(r^{2}+R^{2})}. 
\end{multline}
The second term in equation A6 is
\begin{multline}
-R\int^{r}_{0}{\rm d}r^{\prime}\frac{r^{\prime2}}{(r^{\prime2}+R^{2})^{3}} = \\
\frac{rR}{4(r^{2}+R^{2})^{2}}-\frac{1}{8R^{2}}\left[\arctan\left(\frac{r}{R}\right)+\frac{rR}{r^{2}+R^{2}}\right]. 
\end{multline}
Hence, equation A6 becomes
\begin{multline}
\Omega_{\textsc{g}}(r) = -\uppi^{2}G\rho_{c}^{2}R^{7}\times\left[\frac{2rR}{(r^{2}+R^{2})^{2}}+\right. \\
\left.\frac{1}{R^{2}}\left(\arctan\left[\frac{r}{R}\right]+\frac{rR}{r^{2}+R^{2}}\right)-\frac{4\arctan\left[\frac{r}{R}\right]}{r^{2}+R^{2}}\right]. 
\end{multline}
This is the gravitational potential energy of a truncated $\eta=4$ Plummer-like distribution.

In the limit that $r/R\to\infty$, $\arctan(r/R)\to \frac{\uppi}{2}$, and equation A9 reduces to the gravitational potential energy of an infinite $\eta=4$ Plummer-like distribution,
\begin{equation}
\Omega_{\textsc{g},{\rm inf}}=-\frac{1}{2\uppi}\frac{GM_{\rm inf}^{2}}{R}.
\end{equation}

\section*{APPENDIX B: Gravitational Potential Energy of a Truncated Gaussian Distribution}

\setcounter{equation}{0}
\makeatletter 
\renewcommand{\theequation}{B\@arabic\c@equation}
\makeatother

We give here a brief derivation of the gravitational potential energy of a truncated Gaussian distribution.

We assume a radially-symmetric Gaussian density distribution
\begin{equation}
\rho(r)=\rho_{0}e^{-\nicefrac{r^{2}}{2R^{2}}}, 
\end{equation}
for which the total mass enclosed at radius $r$ is given by
\begin{align}
M(r) & = 4\uppi\rho_{0}\int_{0}^{r}{\rm d}r^{\prime}\,r^{\prime2}\,e^{-r^{\prime2}/{2R^{2}}} \\ 
 & = 4\uppi\rho_{0}\left[R^{3}\sqrt{\frac{\uppi}{2}}\,{\rm erf}\left(\frac{r}{R\sqrt{2}}\right)-R^{2}re^{\nicefrac{-r^{2}}{2R^{2}}}\right], 
\end{align}
and the total mass summed over all radii is given by
\begin{align}
M_{\rm inf} & = 4\uppi\rho_{0}\int_{0}^{\infty}{\rm d}r^{\prime}\,r^{\prime2}\,e^{-r^{\prime 2}/{2R^{2}}} \\ 
 & = 2\sqrt{2}\uppi^{\nicefrac{3}{2}}\rho_{0}R^{3}. 
\end{align}
Using equations A1, B1 and B3, $\Omega_{\textsc{g}}(r)$ is given by
\begin{align}
\Omega_{\textsc{g}}(r) & = -16\uppi^{2}G\rho_{0}^{2}R^{2}\times \notag \\
 &  \quad\int_{0}^{r}{\rm d}r^{\prime}\,r^{\prime}\,e^{\nicefrac{-r^{\prime 2}}{2R^{2}}}\left[R\sqrt{\frac{\uppi}{2}}\,{\rm erf}\left(\frac{r^{\prime}}{R\sqrt{2}}\right)-r^{\prime}e^{\nicefrac{-r^{\prime 2}}{2R^{2}}}\right]. 
\end{align}
The first term in the integral in equation B6 is
\begin{multline}
\int_{0}^{r}{\rm d}r^{\prime}\,r^{\prime}\,e^{\nicefrac{-r^{\prime 2}}{2R^{2}}}R\sqrt{\frac{\uppi}{2}}\,{\rm erf}\left(\frac{r^{\prime}}{R\sqrt{2}}\right) =  \\
R^{3}\sqrt{\frac{\uppi}{2}}\left[\frac{1}{\sqrt{2}}\erf\left(\frac{r}{R}\right)-e^{-\nicefrac{r^{2}}{2R^{2}}}\erf\left(\frac{r}{R\sqrt{2}}\right)\right]. 
\end{multline}
The second term in the integral in equation B6 is 
\begin{equation}
-\int_{0}^{r}{\rm d}r^{\prime}\,r^{\prime 2}e^{\nicefrac{-r^{\prime 2}}{R^{2}}} = R^{3}\left[\frac{1}{2}\frac{r}{R}e^{\nicefrac{-r^{2}}{R^{2}}}-\frac{\sqrt{\uppi}}{4}\erf\left(\frac{r}{R}\right)\right]. 
\end{equation}
Hence, equation B6 becomes
\begin{multline}
\Omega_{\textsc{g}}(r) = -16\uppi^{2}G\rho_{0}^{2}R^{5}\left[\frac{\sqrt{\uppi}}{4}\erf\left(\frac{r}{R}\right)\right. \\
 -\left.\sqrt{\frac{\uppi}{2}}e^{-\nicefrac{r^{2}}{2R^{2}}}\erf\left(\frac{r}{R\sqrt{2}}\right)+\frac{1}{2}\frac{r}{R}e^{-\nicefrac{r^{2}}{R^{2}}}\right]. 
\end{multline}
This is the gravitational potential energy of a truncated Gaussian distribution.  In the limit that $r\to\infty$, $\erf(r/R)\to 1$ and $e^{-(r/R)^{2}}\to 0$, and so equation B9 reduces to the gravitational potential energy of an infinite Gaussian distribution,
\begin{equation}
\Omega_{\textsc{g},{\rm inf}}=-\frac{1}{2\sqrt{\uppi}}\frac{GM_{\rm inf}^{2}}{R}. 
\end{equation}

\FloatBarrier

\section*{APPENDIX C: The Pattle et al. (2015) Virial Plane}

\setcounter{figure}{0}
\makeatletter 
\renewcommand{\thefigure}{C\@arabic\c@figure}
\makeatother

\begin{figure}
\centering
\includegraphics[width=0.47\textwidth]{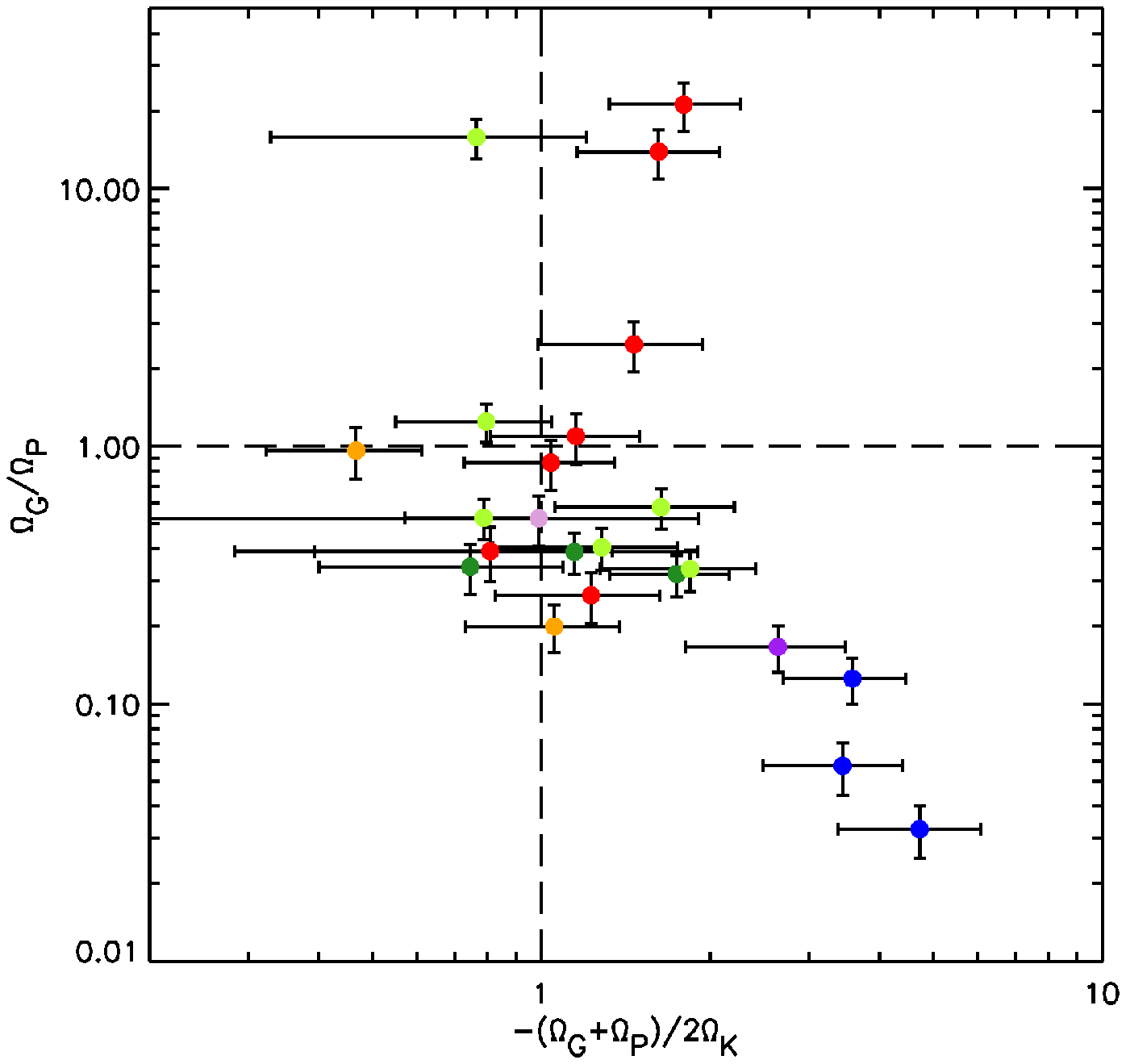}
\caption{The \citet{pattle2015} virial plane for the L1688 region of Ophiuchus.}
\label{fig:2015}
\end{figure}

In Figure~\ref{fig:2015}, we reproduce the virial plane determined by \citet{pattle2015} for the cores in the L1688 region of Ophiuchus, for purposes of comparison with this work.

\label{lastpage}

\end{document}